\documentclass[10pt]{iopart}
\usepackage{epsf,psfig,epsfig} 
\begin{document} 


\newcommand{\bj}{{\bf j}} 
\newcommand{\bk}{{\bf k}} 
\newcommand{\dd}{\mathrm{d}} 

\renewcommand{\bi}{{\bf i}} 
\renewcommand{\lsim}{\stackrel{<}{\sim}}
 
\title{Topological Lensing in Spherical Spaces} 

\author{Evelise Gausmann$^{1}$, Roland Lehoucq$^{2}$, Jean-Pierre
Luminet$^{1}$, Jean-Philippe Uzan$^{3}$ and Jeffrey Weeks$^{4}$}

\address{ (1) D\'epartement d'Astrophysique Relativiste et de Cosmologie,\\
              Observatoire de Paris -- C.N.R.S. UMR 8629, F-92195 Meudon 
              Cedex (France).\\
	      (2) CE-Saclay, DSM/DAPNIA/Service d'Astrophysique,\\
              F-91191 Gif sur Yvette Cedex (France)\\
          (3) Laboratoire de Physique Th\'eorique -- C.N.R.S. UMR 8627, 
              B\^at. 210, Universit\'e
              Paris XI, F-91405 Orsay Cedex (France).\\
          (4) 15 Farmer St., Canton NY 13617-1120, USA.} 
 
\begin{abstract}
	 This article gives the construction and complete classification
	 of all three--dimensional spherical manifolds, and orders them by
	 decreasing volume, in the context of multiconnected universe
	 models with positive spatial curvature.  It discusses which
	 spherical topologies are likely to be detectable by
	 crystallographic methods using three--dimensional catalogs of
	 cosmic objects.  The expected form of the pair separation
	 histogram is predicted (including the location and height of the
	 spikes) and is compared to computer simulations, showing that
	 this method is stable with respect to observational uncertainties
	 and is well suited for detecting spherical topologies.
\end{abstract} 

\pacs{98.80.-q, 04.20.-q, 02.40.Pc} 
 
\section{Introduction}
\label{I} 
 
The search for the topology of our universe has focused mainly on
candidate spacetimes with locally Euclidean or hyperbolic spatial
sections.  This was motivated on the one hand by the mathematical
simplicity of three--dimensional Euclidean manifolds, and on the other
hand by observational data which had, until recently, favored a low
density universe.  More recently, however, a combination of
astrophysical and cosmological observations (among which the
luminosity distance--redshift relation up to $z\sim1$ from type Ia
supernovae~\cite{sndata}, the cosmic microwave background (CMB)
temperature anisotropies~\cite{cmbdata}, gravitational
lensing~\cite{gldata}, velocity fields~\cite{vdata}, and comoving
standard rulers~\cite{srdata}) seems to indicate that the expansion of
the universe is accelerating, with about 70\% of the total energy
density $\Omega_{0}$ being in the form of a dark component
$\Omega_{\Lambda_0}$ with negative pressure, usually identified with a
cosmological constant term or a quintessence field~\cite{quint}.  As a
consequence, the spatial sections of the universe would be ``almost"
locally flat, i.e. their curvature radius would be larger than the
horizon radius ($\sim 10 h^{-1}$ Gpc, where $h$ is the Lema\^\i
tre-Hubble parameter in units of 100 km/s/Mpc).  Recent CMB
measurements~\cite{cmbdata} report a first Doppler peak shifted by a
few percent towards larger angular scales with respect to the peak
predicted by the standard cold dark matter (CDM) inflationary model,
thus favouring a marginally {\it spherical} model~\cite{wsp}.  Indeed,
under specific assumptions such as a $\Lambda$--CDM model and a nearly scale
invariant primordial power spectrum, the value of the total
energy--density parameter is given by $\Omega_{0} \equiv
\Omega_{m_{0}}+\Omega_{\Lambda_{0}} = 1.11^{+0.13}_{-0.12}$ to $95 \%$
confidence~\cite{jaf}.  Note that while flat models still lie well
inside the $95\%$ confidence level, the relation between the angular
diameter distance and the acoustic peak positions in the angular power
spectrum makes the peak positions in models with a low matter content
very dependent on small variations of the cosmological constant.
 
As a consequence of these observable facts, spherical spaceforms are
of increasing interest for relativistic cosmology, in the framework of
Friedmann--Lema\^{\i}tre solutions with positive spatial curvature. 
Due to the current constraints on the spatial curvature of our
universe and to the rigidity theorem~\cite{Mos73}, hyperbolic
topologies may be too large to be detectable by crystallographic
methods.  For example, if $\Omega_{m_{0}} = 0.3$ and
$\Omega_{\Lambda_{0}} = 0.6$, then even in the smallest known
hyperbolic topologies the distance from a source to its nearest
topological image is more than a half of the horizon radius, meaning
that at least one member of each pair of topological images would lie
at a redshift too high to be easily detectable.  When using
statistical methods for detecting a hyperbolic topology, the
topological signature falls to noise level as soon as observational
uncertainties are taken into account.
 
On the other hand, spherical topologies may be easily detectable,
because for a given value of the curvature radius, spherical spaces
can be as small as desired.  There is no lower bound on their volumes,
because in spherical geometry increasing a manifold's complexity
decreases its volume, in contrast to hyperbolic geometry where
increasing the complexity increases the volume.  Thus many different
spherical spaces would fit easily within the horizon radius, no matter
how small $\Omega_{0} - 1$ is.
  
In order to clarify some misleading terminology that is currently used
in the cosmological literature, we emphasize the distinction between
spherical and closed universe models.  A spherical universe has
spatial sections with positive curvature.  The volume of each spatial
section is necessarily finite, but the spacetime can be open (i.e.
infinite in time) if the cosmological constant is high enough.  On the
other hand, a closed universe is a model in which the scale factor
reaches a finite maximum value before recollapsing.  It can be
obtained only if the space sections are spherical {\it and} the
cosmological constant is sufficiently low.
 
As emphasized by many authors (see
e.g.~\cite{lachieze95,uzan97,luminet99,uzan99b} for reviews), the key
idea to detect the topology in a three--dimensional data set is the
{\it topological lens effect}, i.e. the fact that if the spatial
sections of the universe have at least one characteristic size smaller
than the spatial scale of the catalog, then different images of the
same object shoud appear in the survey.  This idea was first
implemented in the {\it crystallographic method}~\cite{lehoucq96},
which uses a pair separation histogram (PSH) depicting the number of
pairs of the catalog's objects having the same three--dimensional
spatial separations in the universal covering space.  Even if
numerical simulations of this method showed the appearance of spikes
related to characteristic distances of the fundamental polyhedron, we
proved that sharp spikes emerge only when the holonomy group has at
least one Clifford translation, i.e. a holonomy that translates all
points the same distance~\cite{lehoucq99} (see also~\cite{gomero98}). 
Since then, various generalisations of the PSH method have been
proposed~\cite{fagundes98,fagundes99,gomero99}
(see~\cite{uzan99b,lehoucq00} for a discussions of these methods) but
none of them is fully satisfactory.  The first step towards such a
generalisation was to exploit the property that, even if there is no
Clifford translation, equal distances in the universal covering space
appear more often than just by chance.  We thus reformulated the
cosmic crystallographic method as a collecting--correlated--pairs
method (CCP)~\cite{uzan99}, where the topological signal was enhanced
by collecting all distance correlations in a single index.  It was
proven that this signal was relevant to detect the topology.

The goal of the present article is twofold.  First we give a
mathematical description of all spherical spaces, which have been
overlooked in the literature on cosmology.  This provides the required
mathematical background for applying statistical methods to detect the
topology: crystallographic methods using three--dimensional data sets
such as galaxy, cluster and quasar catalogs, and methods using
two--dimensional data sets such as temperature fluctuations in the
Cosmic Microwave Background (CMB), and we investigate their
observational signature in catalogs of discrete sources in order to
complete our previous
works~\cite{lehoucq96,lehoucq99,lehoucq00,uzan99} on Euclidean and
hyperbolic manifolds.  We first review in \S~\ref{II} the basics of
cosmology and topology in spherical universes, including the basic
relations to deal with a Friedmann--Lema\^{\i}tre universe of constant
curvature and the basics of cosmic topology.  We then describe and
classify three--dimensional spherical manifolds in \S~\ref{III} and
\S~\ref{III_2} and explain how to use them, with full details gathered
in~\ref{A} and~\ref{B}.  One prediction of our former
works~\cite{lehoucq99,uzan99} was that the PSH method must exhibit
spikes if there exists at least one Clifford translation.  In
\S~\ref{IV} we discuss the spaceforms that are likely to be
detectable.  We show that the location and height of the spikes can be
predicted analytically and then we check our predictions numerically. 
We also study the effect of observational errors on the stability of
the PSH spectra, and briefly discuss the status of the CCP method for
spherical topologies.

\section{Cosmology in a Friedmann--Lema\^{\i}tre spacetime with
spherical spatial sections}
\label{II} 

\subsection{Friedmann--Lema\^{\i}tre spacetimes of constant positive curvature}
 
In this section we describe the cosmology and the basics of the topology of 
universes with spherical spatial sections.  The local geometry of such a 
universe is given by a Friedmann--Lema\^{\i}tre metric 
\begin{equation} 
\dd s^2=-\dd t^2+a^2(t)\left(\dd\chi^2+\sin^2{\chi}\dd\omega^2  
\right).\label{METR} 
\end{equation} 
where $a$ is the scale factor, $t$ the cosmic time and $\dd\omega^2\equiv 
\dd\theta^2+\sin^2{\theta}\dd\varphi^2$ the infinitesimal solid 
angle. $\chi$ is the (dimensionless) comoving radial distance in units 
of the curvature radius $R_C$ of the 3--sphere $S^3$. 

The 3--sphere $S^3$ can be 
embedded in four--dimensional Euclidean space by introducing the set of 
coordinates $(x_\mu)_{\mu=0..3}$ related to the intrinsic coordinates 
$(\chi,\theta,\varphi)$ through (see e.g. ~\cite{wolf83}) 
\begin{eqnarray} 
x_0&=&\cos{\chi}\nonumber\\ 
x_1&=&\sin{\chi}\sin{\theta}\sin{\varphi}\nonumber\\ 
x_2&=&\sin{\chi}\sin{\theta}\cos{\varphi} \nonumber\\ 
x_3&=&\sin{\chi}\cos{\theta}, 
\end{eqnarray} 
with $0\leq\chi\leq\pi$, $0\leq\theta\leq\pi$ and $0\leq\varphi\leq2\pi$. 
The 3--sphere is then the submanifold of equation 
\begin{equation} 
x^\mu x_\mu \equiv x_0^2+x_1^2+x_2^2+x_3^2=+1, 
\end{equation} 
where $x_\mu=\delta_{\mu\nu}x^\nu$.  The comoving spatial distance $d$ 
between any two points $x$ and $y$ on $S^3$ can be computed using the 
inner product $x^\mu y_\mu$.  The value of this inner product is the 
same in all orthonormal coordinate systems, so without loss of 
generality we may assume $x = (1,0,0,0)$ and $y = (\cos d, \sin d, 0, 
0)$, giving $x^\mu y_\mu = \cos d$.  Hence, the comoving spatial 
distance between two points of comoving coordinates $x$ and $y$ is 
given by
\begin{equation} 
d[x,y]={\rm arc}\cos{\left[x^\mu y_\mu\right]}, 
\end{equation}
The volume enclosed by a sphere of radius $\chi$ is, in units of the 
curvature radius,
\begin{equation}
\label{vol_chi} 
\mathrm{Vol}(\chi)=\pi\left(2\chi-\sin{2\chi}\right). 
\end{equation} 

\begin{figure}[ht] 
\centerline{\epsfig{file = 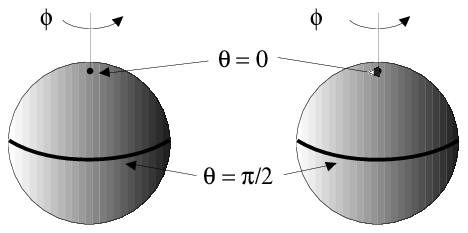, width=9cm}} 
\centerline{\epsfig{file = 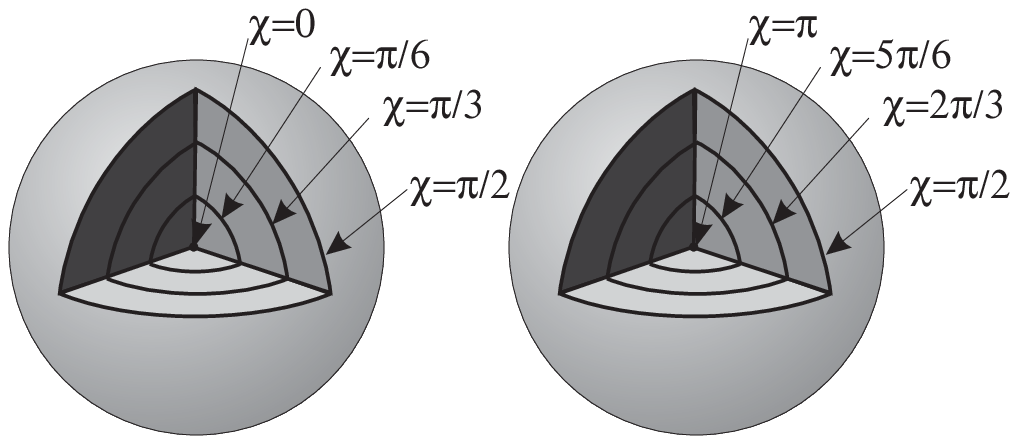, width=9cm}} 
\caption{Representation of $S^3$ by two balls in $R^3$ glued together.
Top: The $\theta$ and $\phi$ coordinates are the standard ones.
Bottom: The $\chi$ coordinate runs from 0 at the center of one ball
(the ``north pole'' of $S^3$) through $\pi/2$ at the ball's surface
(the spherical ``equator'' of $S^3$) to $\pi$ at the center of the
other ball (the ``south pole'' of $S^3$).}
\label{S3} 
\end{figure}

A convenient way to visualize the 3--sphere is to consider $S^3$ as
composed of two solid balls in Euclidean space ${R}^3$, glued
together along their boundaries (figure~\ref{S3}): each point of the
boundary of one ball is the same as the corresponding point in the
other ball.  To represent a point of coordinates $x_\mu$ in a 
three--dimensional space we consider only the coordinates $(x_i)_{i=1..3}$,
which are located in the interior of a ball, and discard the nearly redundant
coordinate $x_0$.  However, the two points of coordinates
$(\chi,\theta,\phi)$ and $(\pi-\chi,\theta,\phi)$, corresponding 
respectively to the points $(x_0, x_1, x_2, x_3)$ and $(-x_0, x_1, x_2, 
x_3)$ in the four--dimensional Euclidean space,  have the same
coordinates ($x_1,x_2,x_3$) and we thus have to use two balls, one
corresponding to $0\leq\chi\leq\pi/2$ (i.e. $x_0 \geq 0$) and the other one to
$\pi/2\leq\chi\leq\pi$ (i.e. $x_0 \leq 0$).  Each ball represents half the space.

All three--dimensional observations provide at least the position of 
an object on the celestial sphere and its redshift $z\equiv a/a_0-1$ 
(the value $a_0$ of the scale factor today may be chosen arbitrarily; 
a natural choice is to set $a_0$ equal to the physical curvature radius 
today). To reconstruct an object's three--dimensional position
$(\chi,\theta,\varphi)$ we need to compute the relation between the 
radial coordinate $\chi$ and the redshift $z$, which requires the law of 
evolution of the scale factor obtained from the Friedmann equations 
\begin{eqnarray} 
H^2&=&\frac{\kappa}{3}\rho-\frac{k}{a^2R_C^2}+\frac{\Lambda}{3}\\ 
\frac{\ddot a}{a}&=&-\frac{\kappa}{6}(\rho+3P)+\frac{\Lambda}{6} 
\end{eqnarray} 
where $\rho$ and $P$ are the energy density and pressure of the cosmic 
fluid, $\Lambda$ the cosmological constant, $\kappa \equiv 8\pi G$ with $G$ the 
Newton constant, $k=+1$ is the curvature index and $H\equiv \dot a/a$ is 
the Hubble parameter. As a first consequence, we deduce from these 
equations that the physical curvature radius today is given by 
\begin{equation}
	\label{rco} 
	R_{C_0}^{\rm phys} \equiv a_{0}R_{C_0} = 
	\frac{c}{H_0}\frac{1}{\sqrt{\left|\Omega_{\Lambda_0} + \Omega_{m_0}-1\right|}} 
\end{equation} 
where the density parameters are defined by 
\begin{equation} 
\Omega_{\Lambda}\equiv\frac{\Lambda}{3H^2} 
\qquad 
\Omega_{m}\equiv\frac{\kappa\rho}{3H^2}. 
\end{equation} 
As emphasized above, we can choose $a_0$ to be the physical curvature 
radius today, i.e. $a_0 = R_{C_0}^{\rm phys}$, which 
amounts to choosing the units on the comoving sphere such that $R_{C_0} 
=1$, hence determining the value of the constant $a_0$. 
As long as we are dealing with catalogs of galaxies or clusters, we can 
assume that the universe is filled with a pressureless 
fluid. Integrating the radial null geodesic equation $\dd\chi=\dd t/a$ 
leads  to 
\begin{equation}
\chi(z)=\int_0^z\frac{\sqrt{\Omega_{m_0}+\Omega_{\Lambda_0}-1}\dd x} 
{\sqrt{\Omega_{\Lambda_0}+(1-\Omega_{m_0}-\Omega_{\Lambda_0})(1+x)^2 
+\Omega_{m_0}(1+x)^3}}.\label{chi2} 
\end{equation}

\subsection{Basics of cosmic topology}
 
Equations (\ref{METR})--(\ref{chi2}) give the main properties that
describe the local geometry, i.e. the geometry of the universal
covering space $\Sigma$, independently of the topology.  Indeed it is
usually assumed that space is simply connected so that the spatial
sections are the 3--sphere $S^3$.  To describe the topology of these
spatial sections we have to introduce some basic topological elements
(see \cite{lachieze95} for a review).  From a topological point of
view, it is convenient to describe a three--dimensional
multi--connected manifold $M$ by its fundamental polyhedron, which is
convex with an even number of faces.  The faces are identified by
face--pairing isometries.  The face--pairing isometries generate the
{holonomy group} ${\Gamma}$, which acts without fixed points on the
three--dimensional covering space $\Sigma$ (see
\cite{wolf83,beardon83,nakahara90} for mathematical definitions and
\cite{lachieze95,uzan99b} for an introduction to topology in the
cosmological context).  The holonomy group $\Gamma$ is isomorphic to
the first fundamental group $\pi_{1}(M)$.

\begin{figure}[ht] 
\centerline{\epsfig{file = 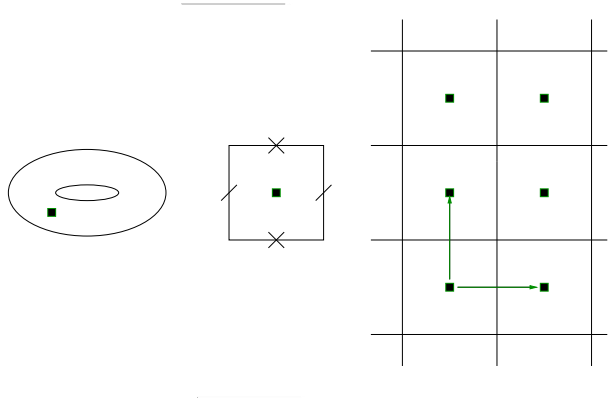,width=8cm}}
\caption{Illustration of the general topological definitions in the case
of a two--dimensional torus (left). Its fundamental polyhedron is a square,
opposite faces of which are identified (middle).
In its universal covering space (right) the two translations taking
the square to its nearest images generate the holonomy group.}
\label{tore} 
\end{figure}

To illustrate briefly these definitions, let us consider the
particularly simple case of a two--dimensional flat torus $T^2$.  It
can be constructed from a square, opposite edges of which are glued
together.  The translations taking one edge to the other are the face
pairing isometries.  The holonomy group of this space, generated by
the face--pairing translations, is isomorphic to the group of loops on
the torus $\pi_{1}(T^2)=Z\times Z$ (see figure~\ref{tore}).

In our case $\Sigma=S^3$ and its isometry group is the rotation group of
the four--dimensional space in which it sits, i.e. $G=SO(4)$.  In units
of the curvature radius, the volume of the spatial sections is given by
\begin{equation} 
\hbox{Vol}(S^3/\Gamma)=\frac{\hbox{Vol}(S^3)}{|\Gamma|}=\frac{2\pi^2} 
{|\Gamma|},\label{vol_ell} 
\end{equation} 
where $|\Gamma|$ is the order of the group $\Gamma$, i.e. the number 
of elements contained in $\Gamma$.
Since the elements $g$ of $\Gamma$ are isometries, they satisfy 
\begin{eqnarray} 
\forall x,y\in\Sigma\quad \forall g\in\Gamma,\quad 
\hbox{dist}[x,y]=\hbox{dist}[g(x),g(y)] 
\end{eqnarray} 
An element $g$ of $\Gamma$ is a Clifford translation if it translates all points the  
same distance, i.e. if  
\begin{eqnarray} 
\forall x,y\in\Sigma\quad  
\hbox{dist}[x,g(x)]=\hbox{dist}[y,g(y)],
\label{Clifford}
\end{eqnarray}  
The significance of these particular holonomies, which are central to
the detection of the topology, will be explained in section
\ref{generalities}.  To perform computations we express the isometries
of the holonomy group $\Gamma \subset SO(4)$ as $4\times4$ matrices. 
To enumerate all spherical manifolds we will need a classification of
all finite, fixed-point free subgroups of $SO(4)$.  This will be the
purpose of sections \ref{III} and \ref{III_2}.

\subsection{Spherical spaces and cosmology} 
 
As early as 1917, de Sitter \cite{deS17} distinguished the sphere
$S^3$ from the projective space $P^3$ (which he called elliptical
space) in a cosmological context.  Both spaceforms are finite, with
(comoving) volumes $2 \pi ^2$ and $ \pi ^2$, respectively.  The
projective space $P^3$ is constructed from the sphere $S^3$ by
identifying all pairs of antipodal points.  The main difference
between them is that in a sphere all straight lines starting from a
given point reconverge at the antipodal point, whereas in projective
space two straight lines can have at most one point in common.  Thus
the sphere does not satisfy Euclid's first axiom, while projective
space does.  In $S^3$ the maximal distance between any two points is
$\pi$, and from any given point there is only one point, the antipodal
one, at that maximal distance.  In $P^3$ the maximal distance is $
\pi/2$ and the set of points lying at maximal distance from a given
point forms a two--dimensional projective plane $P^2$.  Because each
pair of antipodal point points in the 3--sphere projects to a single
point in projective space, if we adopt the 3--sphere as our space
model we may be representing the physical world in duplicate.  For
that reason, de Sitter claimed that $P^3$ was really the simplest
case, and that it was preferable to adopt this in a cosmological
context.  Einstein did not share this opinion, and argued that the
simple connectivity of $S^3$ was physically preferable \cite{Lum97}. 
Eddington \cite{Edd23}, Friedmann \cite{Fri24} and Lema\^\i tre
\cite{Lem29} also referred to projective space as a more physical
alternative to $S^3$.

Narlikar and Seshadri \cite{Nar85} examined the conditions under which
ghost images of celestial objects may be visible in a
Friedmann--Lema\^{\i}tre model with the topology of projective space. 
De Sitter had correctly remarked that the most remote points probably
lie beyond the horizon, so that the antipodal point, if any, would
remain unobservable.  Indeed, given the present data, it will be
impossible to distinguish $S^3$ from $P^3$ observationally if
$\Omega_{0} -1 \ll 1$, because our horizon radius is too small.

None of these authors mentioned other multi--connected spaces in a
cosmological context.  This was first done by Ellis \cite{Ell71} and
Gott \cite{Got80}.  More recently, lens spaces have been investigated
in the framework of quantum gravity models~\cite{Ioni98} and in terms
of their detectability \cite{gom01}.  Nevertheless, the literature on
multi--connected spherical cosmologies is underdeveloped; here we aim
to fill the gap.

\section{The Mathematics of Spherical Spaces I: Classification of $S^3$ subgroups}
\label{III} 
 
This section describes the classification of all spherical 3--manifolds 
and develops an intuitive understanding of their topology and geometry. 
Of particular observational relevance, we will see which spherical 
3--manifolds have Clifford translations in their holonomy groups and 
which do not.  Threlfall and Seifert~\cite{ThrelfallSeifert} gave the 
first complete classification of these spherical 3--manifolds.  Our approach 
borrows heavily from theirs, while also making use of quaternions as 
in~\cite{thurston97}.

As emphasized in the previous section, technically what we will need in
order to perform any computation is the form of the holonomy
transformations $g$ as $4\times 4$ matrices in $SO(4)$, and an
enumeration of all finite subgroups $\Gamma \subset SO(4)$.  We will
enumerate the finite subgroups of $SO(4)$ (see
sections~\ref{generalities} and \ref{III_2}) in terms of the simpler
enumeration of finite subgroups of $SO(3)$ (see section~\ref{SO3}).  The
connection between $SO(4)$ and $SO(3)$ will use quaternions (see \ref{A}
and \ref{B}).

\subsection{Generalities}
\label{generalities}

Our starting point is the fact that for each isometry $g\in O(n+1)$ 
there is a basis of $R^{n+1}$ relative to which the matrix of $g$ 
has the form 
\begin{equation}
\left( 
\begin{array}{ccccccrrc} 
  +1     &  0     & \cdots &     &        &    &             &              &        \\ 
   0     & \ddots &        &     &        &    &             &              &        \\ 
         &        & +1     &     &        &    &             &              &        \\ 
  \vdots &        &        &  -1 &        &    &             &              &        \\ 
         &        &        &     & \ddots &    &             &              &        \\ 
         &        &        &     &        & -1 &             &              &        \\ 
         &        &        &     &        &    & \cos \alpha & -\sin \alpha &        \\ 
         &        &        &     &        &    & \sin \alpha &  \cos \alpha &        \\ 
         &        &        &     &        &    &             &              & \ddots \\ 
\end{array} 
\right) 
\end{equation} 
If $g$ is a holonomy transformation of an $n$--manifold, 
then $g$ has no fixed points and its matrix has no $+1$ terms 
on the diagonal.  Furthermore, each pair of $-1$ terms may be rewritten 
as a sine--cosine block with $\alpha=\pi$.  Thus when $n=3$ 
the matrix takes the form 
\begin{equation}\label{CANONICAL_MATRIX} 
M(\theta, \phi)= 
\left( 
\begin{array}{cccc} 
      \cos \theta  &  -\sin \theta  &     0    &      0    \\ 
      \sin \theta  &   \cos \theta  &     0    &      0    \\ 
          0       &       0       & \cos \phi & -\sin \phi \\ 
          0       &       0       & \sin \phi &  \cos \phi \\ 
\end{array}\right) 
\end{equation} 
An immediate consequence of this decomposition is that every spherical
3--manifold is orientable.  Indeed all odd-dimensional spherical
manifolds must be orientable for this same reason.  In even dimensions
the only spherical manifolds are the $n$--sphere $S^n$ (which is
orientable) and the $n$--dimensional projective space $P^n$ (which
is non orientable).
 
If we replace the static matrix $M(\theta, \phi)$ with the
time--dependent matrix $M(\theta t, \phi t)$, we generate a flow on
$S^3$, i.e.
 to each point $x\in S^3$ we associate the flow line
$x(t)=M(\theta t, \phi t)x$.  This flow is most beautiful in the
special case $\theta = \pm \phi$.  In this special case \emph{all}
flow lines are geodesics (great circles), and the flow is homogeneous
in the sense that there is an isometry of $S^3$ taking any flow line
to any other flow line.  The matrix $M(\theta, \pm \theta)$ defines a
\emph{Clifford translation} because it translates all points the same
distance (see equation~\ref{Clifford}).  We further distinguish two
families of Clifford translations.  When $\theta= \phi$ the flow lines
spiral clockwise around one another, and the Clifford translation is
considered \emph{right--handed} whereas when $\theta = -\phi$ the flow
lines spiral anticlockwise around one another, and the Clifford
translation is considered \emph{left--handed}.
 
Every isometry $M(\theta, \phi) \in SO(4)$ is the product 
of a right--handed Clifford translation $M(\alpha,\alpha)$ 
and a left--handed Clifford translation $M(\beta,-\beta)$ as 
\begin{eqnarray} 
M(\theta,\phi) = M(\alpha,\alpha)\, M(\beta,-\beta) =
M(\beta,-\beta)\, M(\alpha,\alpha)
\end{eqnarray} 
where $\alpha\equiv(\theta + \phi)/2$ and $\beta\equiv(\theta -
\phi)/2$ and the order of the factors makes no difference.  This
factorization is unique up to simultaneously multiplying both factors
by -1.  Moreover every right--handed Clifford translation commutes
with every left--handed one, because there is always a coordinate
system that simultaneously brings both into their canonical form
(\ref{CANONICAL_MATRIX}).
 
Just as the unit circle $S^1$ enjoys a group structure as the set
${\cal S}^1$ of complex numbers of unit length, the 3--sphere $S^3$
enjoys a group structure as the set ${\cal S}^3$ of quaternions of
unit length (see~\ref{A} for details).  Each right--handed Clifford
translation corresponds to left multiplication by a unit length
quaternion (${\bf q} \rightarrow {\bf x}{\bf q}$), so the group of all
right--handed Clifford translations is isomorphic to the group ${\cal
S}^3$ of unit length quaternions, and similarly for the left--handed
Clifford translations, which correspond to right multiplication (${\bf
q}\rightarrow{\bf q}{\bf x}$). It follows that $SO(4)$ is isomorphic
to ${\cal S}^3\times{\cal S}^3/\lbrace\pm({\bf 1},{\bf 1})\rbrace$, 
where {\bf 1} is the identity quaternion, 
so the classification of all subgroups of $SO(4)$ can be deduced
from the classification of all subgroups of ${\cal S}^3$.

Classifying all finite subgroups of ${\cal S}^3$ seems difficult at
first, but luckily it reduces to a simpler problem.  The key is to
consider the action of the quaternions by conjugation.  That is, for
each unit length quaternion ${\bf x} \in {\cal S}^3$, consider the
isometry $p_{\bf x}$ that sends each quaternion ${\bf q}$ to ${\bf x}
{\bf q} {\bf x}^{-1}$
\begin{equation} 
p_{\bf x}: \left\lbrace 
\begin{array}{l} 
{\cal S}^3 \rightarrow {\cal S}^3\\ 
{\bf q}\longmapsto p_{\bf x}({\bf q})={\bf x} {\bf q} {\bf 
x}^{-1} 
\end{array}\right.. 
\end{equation} 
The isometry $p_{\bf x}$ fixes the identity quaternion {\bf 1}, so in
effect its action is confined to the equatorial 2--sphere spanned by
the remaining basis quaternions (\bi, \bj, \bk) [see~\ref{A} for
details and definitions concerning quaternions].  Thus, by
restricting our attention to the equatorial 2--sphere, we get an
isometry
\begin{equation} 
p_{\bf x}: S^2\rightarrow S^2 
\end{equation} 
In other words, each ${\bf x}\in {\cal S}^3$ defines an element
$p_{\bf x}\in SO(3)$, and the mapping
\begin{equation} 
p: \left\lbrace 
\begin{array}{l} 
{\cal S}^3 \rightarrow SO(3)\\ 
{\bf x}\longmapsto p({\bf x})=p_{\bf x} 
\end{array}\right.
\end{equation}
is a {\it homomorphism} from ${\cal S}^3$ to $SO(3)$. To classify
all subgroups of ${\cal S}^3$ we must first know
the finite subgroups of $SO(3)$.
 
\subsection{Finite Subgroups of $SO(3)$}
\label{SO3}
 
The finite subgroups of $SO(3)$ are just the finite rotation groups 
of a 2--sphere, which are known to be precisely the following: 
\begin{itemize} 
    \item    The {\it cyclic groups} $Z_n$ of order $n$, generated by 
                a rotation through an angle ${2 \pi}/{n}$ about some axis. 
 
    \item    The {\it dihedral groups} $D_m$ of order $2m$, generated by 
                a rotation through an angle ${2 \pi}/{m}$ about some axis 
                as well as a half turn about some perpendicular axis. 
 
    \item    The {\it tetrahedral group} $T$ of order 12 consisting of all 
                orientation--preserving symmetries of a regular tetrahedron. 
 
    \item    The {\it octahedral group} $O$ of order 24 consisting of all 
                orientation--preserving symmetries of a regular octahedron. 
 
    \item    The {\it icosahedral group} $I$ of order 60 consisting of all 
                orientation--preserving symmetries of a regular icosahedron. 
\end{itemize}

If the homomorphism $p: {\cal S}^3 \rightarrow SO(3)$ were an isomorphism, the 
above list would give the finite subgroups of ${\cal S}^3$ directly.  We are 
not quite that lucky, but almost: the homomorphism $p$ is two--to--one. 
It is easy to see that $p_{\bf x} = p_{-\bf x}$ because ${\bf x} {\bf q} 
{\bf x}^{-1} = (-{\bf x}) {\bf q} (-{\bf x})^{-1}$ for all ${\bf q}$. 
There are no other redundancies, so the kernel of $p$ is  
\begin{equation} 
\hbox{Ker}(p)=\lbrace \pm {\bf 1}\rbrace. 
\end{equation} 
 
Let $\Gamma$ be a finite subgroup of ${\cal S}^3$ and consider separately the 
cases that $\Gamma$ does or does not contain $-{\bf 1}$. 

\begin{itemize} 
\item{\it Case 1}: If $\Gamma$ does not contain $-{\bf 1}$, then $p$
maps $\Gamma$ one--to--one onto its image in $SO(3)$, and $\Gamma$ is
isomorphic to one of the groups in the above list.  Moreover, because
we have excluded $-{\bf 1}$, and ${\cal S}^3$ contains no other
elements of order 2, we know that $\Gamma$ contains no elements of
order 2.  The only groups on the above list without elements of order
2 are the cyclic groups $Z_n$ of odd order.  Thus only cyclic groups
of odd order may map isomorphically from ${\cal S}^3$ into $SO(3)$,
and it is easy to check that they all do, for example by choosing the
generator of $Z_n$ to be the quaternion $\cos({2 \pi}/{n}) ~ {\bf 1} +
\sin({2 \pi}/{n}) ~ \bi$.
 
\item{\it Case 2}: If $\Gamma$ contains $-{\bf 1}$, then $p$ maps
$\Gamma$ two--to--one onto its image in $SO(3)$, and $\Gamma$ is a
two--fold cover of one of the groups in the above list.  Conversely,
every group on the list lifts to a group $\Gamma \subset {\cal S}^3$. 
The construction of the two--fold cover is trivially easy: just take
the preimage of the group under the action of $p$.  The result is
called the binary cyclic, binary dihedral, binary tetrahedral, binary
octahedral, or binary icosahedral group.  A ``binary cyclic group'' is
just a cyclic group of twice the order, so all even order cyclic
groups occur in this fashion.  The remaining binary groups are
\emph{not} merely the product of the original polyhedral group with a
$Z_2$ factor, nor are they isomorphic to the so--called extended
groups which include the orientation--reversing as well as the
orientation-preserving symmetries of the given polyhedron, but are
something completely new\footnote{The binary dihedral group
$D_{1}^{*}$ is isomorphic to the plain cyclic group $Z_{4}$.}.  Note
that the plain dihedral, tetrahedral, octahedral, and icosahedral
groups do not occur as subgroups as of ${\cal S}^3$ -- only their
binary covers do.
\end{itemize} 

\subsection{Finite Subgroups of ${\cal S}^3$} 

Combining the results of the two previous cases, we get the complete 
classification of finite subgroups of the group ${\cal S}^3$ of unit 
length quaternions as follows:

\begin{itemize} 
    \item    The cyclic groups $Z_n$ of order $n$. 
 
    \item    The binary dihedral groups $D_m^{\ast}$ of order $4m$, $m \ge 2$. 
 
    \item    The binary tetrahedral group $T^{\ast}$ of order 24. 
 
    \item    The binary octahedral group $O^{\ast}$ of order 48. 
 
    \item    The binary icosahedral group $I^{\ast}$ of order 120. 
\end{itemize} 
 
\section{The Mathematics of Spherical Spaces II: Classification of 
spherical spaceforms}\label{III_2} 

There are three categories of spherical 3--manifolds.  The
\emph{single action manifolds} are those for which a subgroup $R$ of
${\cal S}^3$ acts as pure right--handed Clifford translations.  The
\emph{double action manifolds} are those for which subgroups $R$ and
$L$ of ${\cal S}^3$ act simultaneously as right-- and left--handed
Clifford translations, and every element of $R$ occurs with every
element of $L$.  The \emph{linked action manifolds} are similar to the
double action manifolds, except that each element of $R$ occurs with
only some of the elements of $L$.

After introducing some definitions, we give the classifications
of single action manifolds (\S~\ref{sa}), double action
manifolds (\S~\ref{da}) and linked action manifolds (\S~\ref{la}) and
present in \S~\ref{sum} a summary of these classifications.

\begin{figure}[ht] 
\centerline{\epsfig{file=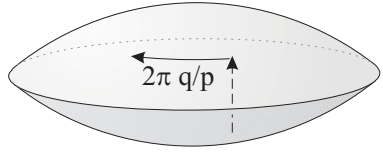, width=5cm}} 
\caption{Construction of a lens space $L(p,q)$.}
\label{figlens} 
\end{figure}

We define a lens space $L(p,q)$ by identifying the lower surface of a
lens-shaped solid to the upper surface with a ${2\pi}q/p$ rotation
(see figure \ref{figlens}), for relatively prime integers $p$ and $q$
with $0 < q < p$.  Furthermore, we may restrict our attention to $0 <
q \leq p/2$ because for values of $q$ in the range $p/2 < q < p$ the
twist ${2\pi}q/p$ is the same as $-{2\pi}(p-q)/p$, thus $L(p,q)$ is
the mirror image of $L(p,p-q)$.  When the lens is drawn in Euclidean
space its faces are convex, but when it is realized in the 3--sphere
its faces lie on great 2--spheres, filling a hemisphere of each. 
Exactly $p$ copies of the lens tile the universal cover $S^3$, just as
the 2--dimensional surface of an orange may be tiled with $p$ sections
of orange peel, each of which is a bigon with straight sides meeting
at the poles.  Two lens spaces $L(p,q)$ and $L(p',q')$ are
homeomorphic if and only if $p=p'$ and either $q=\pm q' ({\rm
mod}\,p)$ or $qq' = \pm 1 ({\rm mod}\,p)$.

A cyclic group $Z_n$ may have several different realizations as
holonomy groups $\Gamma \subset SO(4)$.  For example, the lens spaces
$L(5,1)$ and $L(5,2)$ are nonhomeomorphic manifolds, even though their
holonomy groups are both isomorphic to $Z_5$.  For noncyclic groups,
the realization as a holonomy group $\Gamma \subset SO(4)$ is unique
up to an orthonormal change of basis, and thus the resulting manifold
is unique.

\subsection{Single Action Spherical 3-Manifolds}\label{sa} 

The finite subgroups of ${\cal S}^3$ give the single action manifolds
directly, which are thus the simplest class of spherical 3--manifolds.
They are all given as follows:
 
\begin{itemize} 
	\item Each cyclic group $Z_n$ gives a lens space $L(n,1)$, whose
	fundamental domain is a lens shaped solid, $n$ of which tile the 
	3--sphere.
 
	\item Each binary dihedral group $D_m^{\ast}$ gives a prism
	manifold, whose fundamental domain is a $2m$--sided prism, $4m$ of
	which tile the 3--sphere.
 
	\item The binary tetrahedral group $T^{\ast}$ gives the octahedral
	space, whose fundamental domain is a regular octahedron, 24 of
	which tile the 3--sphere in the pattern of a regular 24--cell.
 
	\item The binary octahedral group $O^{\ast}$ gives the truncated
	cube space, whose fun\-da\-men\-tal domain is a truncated cube, 48
	of which tile the 3--sphere.
 
	\item The binary icosahedral group $I^{\ast}$ gives the Poincar\'e
	dodecahedral space, whose fundamental domain is a regular
	dodecahedron, 120 of which tile the 3--sphere in the pattern of a
	regular 120--cell.  Poincar\'e discovered this manifold in a
	purely topological context, as the first example of a multiply
	connected homology sphere~\cite{Poincare04}.  A quarter century
	later Weber and Seifert glued opposite faces of a dodecahedron and
	showed that the resulting manifold was homeomorphic to
	Poincar\'e's homology sphere~\cite{WeberSeifert33}.
\end{itemize} 

\begin{figure}[ht] 
\centerline{\epsfig{file=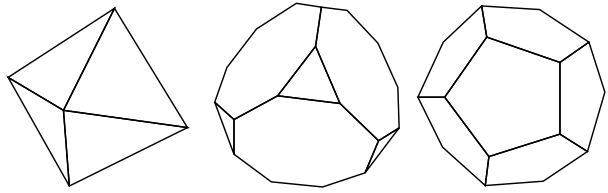, width=8cm}} 
\caption{Fundamental domains for three single action 3--manifolds. 
From left to right, the regular octahedron, the truncated cube and the
regular dodecahedron which respectively correspond to the spaces
generated by the binary tetrahedral group $T^*$, the binary octahedral
group $O^*$ and the binary icosahedral group $I^{\ast}$.}
\label{TOI} 
\end{figure}

\begin{figure}[ht] 
\centerline{\epsfig{file=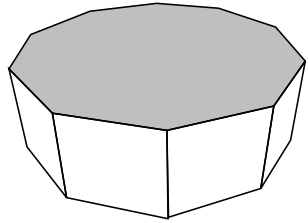, width=4cm}} 
\caption{An example of fundamental domain for a prism manifold. This
10--sided prism is the fundamental polyhedron of the space generated by
the binary dihedral group $D_5^*$, of order 20.}
\label{fig_prism} 
\end{figure}

In figure~\ref{TOI}, we depict the fundamental domains for the binary
tetrahedral group $T^*$, binary octahedral group $O^*$ and binary
icosahedral group $I^{\ast}$.  The fundamental polyhedron for the lens
space $L(n,1)$ can be constructed by following the example presented
in figure~\ref{figlens}.  Finally, the fundamental domain of the prism
manifold generated by the binary dihedral group $D_5^*$ is shown in
figure~\ref{fig_prism}.

To finish, let us emphasize that all single action manifolds are
globally ho\-mo\-ge\-neous, in the sense that there is an isometry $h$ of
the manifold taking any point $x$ to any other point $y$.  If the
manifold's holonomy group is realized as left multiplication by a
group $\Gamma = \lbrace g_i \rbrace$ of quaternions, then the isometry
$h$ is realized as right multiplication by $x^{-1} y$.  To check that
$h$ is well--defined on the quotient manifold $S^3/\Gamma$, note that
$h$ takes any point $g_i x$ equivalent to $x$ to a point $g_i x
(x^{-1} y) = g_i y$ equivalent to $y$, thus respecting equivalence
classes of points.

\subsection{Double action spherical 3-manifolds}\label{da}
 
The double action spherical 3--manifolds are obtained by letting one
finite subgroup $R \subset {\cal S}^3$ act as right--handed Clifford
translations (equivalent to left multiplication of quaternions) while
a different finite subgroup $L \subset {\cal S}^3$ simultaneously acts
as left--handed Clifford translations (equivalent to right
multiplication of quaternions).  A priori any two subgroups of ${\cal
S}^3$ could be used.  However, if an element $M(\theta, \theta)$ of
$R$ and an element $M(\phi, -\phi)$ of $L$ share the same translation
distance $ | \theta | = | \phi | $, then their composition, which
equals $M(2 \theta, 0)$ or $M(0, 2 \theta)$, will have fixed points
unless $\theta = \phi = \pi$.  Allowing fixed points would take us
into the realm of orbifolds, which most cosmologists consider
unphysical to describe our universe, so we do not consider them here. 
In practice this means that the groups $R$ and $L$ cannot contain
elements of the same order, with the possible exception of $\pm {\bf
1}$.  The binary dihedral, tetrahedral, octahedral, and icosahedral
groups all contain elements of order 4, and so cannot be paired with
one another.

Thus either $R$ or $L$ must be cyclic.  Without loss of generality we
may assume that $L$ is cyclic.  If $R$ is also cyclic, then $L$ and
$R$ cannot both contain elements of order 4, and so we may assume that
it is $L$ that has no order 4 elements.  If $R$ is not cyclic, then
$R$ is a binary polyhedral group, and again $L$ can have no elements
of order 4.  Thus either $L = Z_n$ or $L = Z_{2n}$, with $n$ odd. 
Recalling that the group $Z_n$ consists of the powers $ \lbrace q^i
\rbrace _{0 \leq i < n}$ of a quaternion $q$ of order $n$, it's
convenient to think of the group $Z_{2n}$ as $ \lbrace q^i \rbrace _{0
\leq i < n} \cup \lbrace -q^i \rbrace _{0 \leq i < n}$.  If $R$
contains $-{{\bf 1}}$, then nothing is gained by including the
$\lbrace -q^i \rbrace$ in $L$, because each possible element $(r)(-l)$
already occurs as $(-r)(l)$.  In other words, $L = Z_n$ and $L =
Z_{2n}$ produce the same result, the only difference being that with
$L = Z_{2n}$ each element of the resulting group is generated twice,
once as $(r)(l)$ and once as $(-r)(-l)$.  If $R$ does not contain
$-{{\bf 1}}$, then it is cyclic of odd order and we swap the roles of
$R$ and $L$.  Either way, we may assume $L$ is cyclic of odd order. 
The double action spherical 3--manifolds are therefore the following:

\begin{itemize} 
    \item    $R = Z_m$ and $L = Z_n$, with $m$ and $n$ relatively prime,
                always yields a lens space $L(m n, q)$. 
                 However, not all lens spaces arise in this way.
 
    \item    $R = D_m^{\ast}$ and $L = Z_n$, with $\gcd(4m,n) = 1$, 
                yields an $n$--fold quotient of a prism manifold 
                that is simultaneously a $4m$--fold quotient of 
                the lens space $L(n,1)$. 
 
    \item    $R = T^{\ast}$ and $L = Z_n$, 
                with $\gcd(24,n) = 1$, yields an $n$--fold quotient 
                of the octahedral space that is simultaneously 
                a 24--fold quotient of the lens space $L(n,1)$. 
 
    \item    $R = O^{\ast}$ and $L = Z_n$, 
                with $\gcd(48,n) = 1$, yields an $n$--fold quotient 
                of the truncated cube space that is simultaneously 
                a 48--fold quotient of the lens space $L(n,1)$. 
 
    \item    $R = I^{\ast}$ and $L = Z_n$, 
                with $\gcd(120,n) = 1$, yields an $n$--fold quotient 
                of the Poincar\'e dodecahedral space 
                that is simultaneously a 120--fold quotient of 
                the lens space $L(n,1)$. 
\end{itemize} 

\begin{figure}[ht] 
\centerline{\epsfig{file=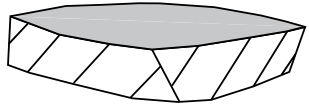, width=4cm}} 
\caption{A fundamental domain for the double action manifold of order
240 generated by the binary octahedral group $R=O^*$ and the cyclic
group $L=Z_5$.}
\label{jeff2} 
\end{figure}

In figure~\ref{jeff2}, we present the fundamental domain of the
double action manifold generated by the binary octahedral group
$R=O^*$ and the cyclic group $L=Z_5$, of orders 48 and 5 respectively.

\subsection{Linked Action Spherical 3--Manifolds}
\label{la}
 
The third and final way to construct spherical 3--manifolds is 
to choose groups $R$ and $L$ as before, but allow each element 
$r \in R$ to pair with a restricted class of elements $l \in L$, 
being careful to exclude combinations of $r$ and $l$ 
that would create fixed points. 
The following \emph{linked action manifolds} arise in this way. 

\begin{itemize} 
   \item $R$ and $L$ are the cyclic groups respectively generated by
         $$ r = M\left(\frac{p+q+1}{2 p} 2\pi, \,
            \frac{p+q+1}{2 p} 2\pi \right)$$ 
         and 
         $$l = M\left(\frac{p-q+1}{2 p} 2\pi, \,
               -\frac{p-q+1}{2 p} 2\pi \right)
         $$
         with $0 < q < p$ and $\gcd(p,q) = 1$.  The generator $r$ is
	     linked to the generator $l$, and their powers are linked
	     accordingly. This yields the lens space $L(p,q)$, generated by
	     $r l = M\left(2\pi/p, 2\pi q/p \right)$.
	     Note that when $p + q$ is even, each element of $L(p,q)$
	     is produced twice, once as $r^k l^k$ and once as
	     $r^{p+k} l^{p+k} = r^p r^k l^p l^k = (-r^k)(-l^k)$.

	\item $R = T^{\ast}$ and $L = Z_{9n}$ with $n$ odd.  The plain
	(not binary) tetrahedral group $T$ contains a normal subgroup $H
	\simeq D_2$ consisting of the three half turns plus the identity. 
	Each element $r$ in the binary tetrahedral group $T^{\ast}$ is
	assigned an index 0, 1, or 2 according to the coset of $H$ in
	which its projection $p_r \in SO(3)$ lies.  Each element $l$ in
	$Z_{9n}$ is assigned an index 0, 1, or 2 equal to its residue
	modulo 3.  An element $r$ is linked to an element $l$ if and only
	if their indices are equal.  This yields a holonomy group with
	only a third as many elements as the full double action group
	would have, and avoids elements with fixed points.
     
	\item $R = D_m^{\ast}$ and $L = Z_{8n}$ with $\gcd(m,8n) = 1$. 
	Each element $r$ in $D_m^{\ast}$ is assigned an index 0 or 1
	according to whether its projection $p_r \in SO(3)$ lies in the
	cyclic part of the plain $D_m$ or not.  Each element $l$ in
	$Z_{8n}$ is assigned an index 0 or 1 equal to its residue modulo
	2.  An element $r$ is linked to an element $l$ if and only if
	their indices are equal.  This yields a holonomy group with only
	half as many elements as the full double action group would have,
	and avoids elements with fixed points.  Note that because $R$ and
	$L$ both contain $-{{\bf 1}}$, each element in the group is
	produced twice, once as $r l$ and once as $(-r) (-l)$.
\end{itemize} 

\begin{figure}[ht] 
\centerline{\epsfig{file=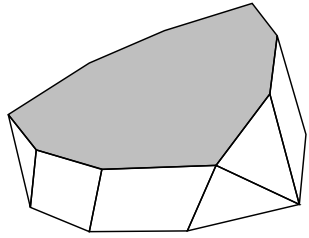, width=4cm}} 
\caption{Fundamental domain of the linked action manifold of order 72 generated by
the binary tetrahedral group $R=T^*$ and the cyclic group $L=Z_9$.}
\label{oz5} 
\end{figure}

As an example, figure~\ref{oz5} shows the fundamental
polyhedron of the linked action manifold generated by the binary
tetrahedral group $R=T^*$ and the cyclic group $L=Z_9$, of orders
24 and 9 respectively.

\subsection{Summary of classification}\label{sum}
 
The preceding sections constructed all three--dimensional spherical
manifolds and classified them in three families. To sum up,
figure~\ref{classif} organizes these manifolds by decreasing volume. 
Since none homeomorphic lens spaces can have isomorphic holonomy 
group, figure~\ref{classif2} charts this special case of lens spaces.\\

\begin{figure}[ht] 
\centerline{\epsfig{file=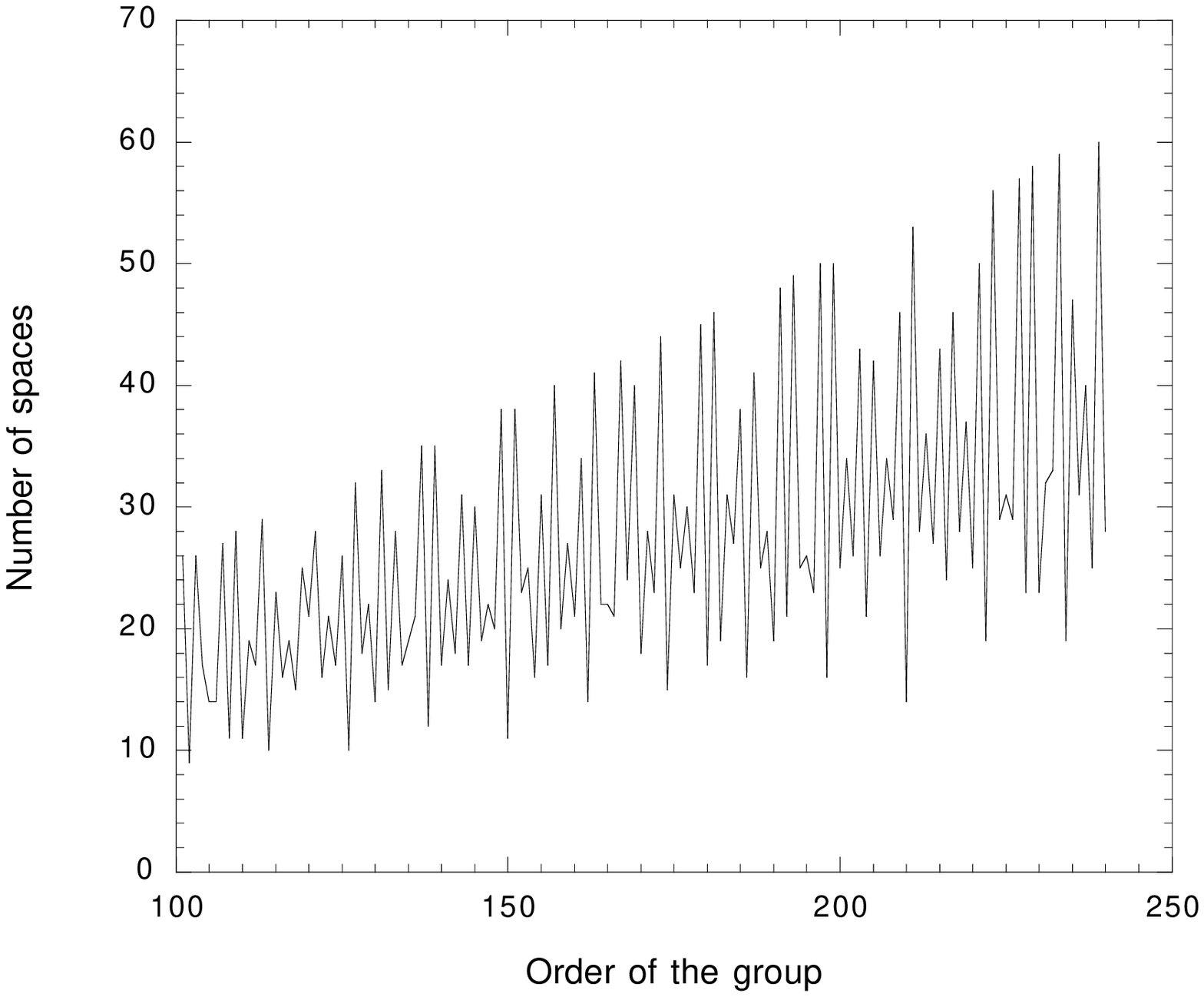, width=7cm},
            \epsfig{file=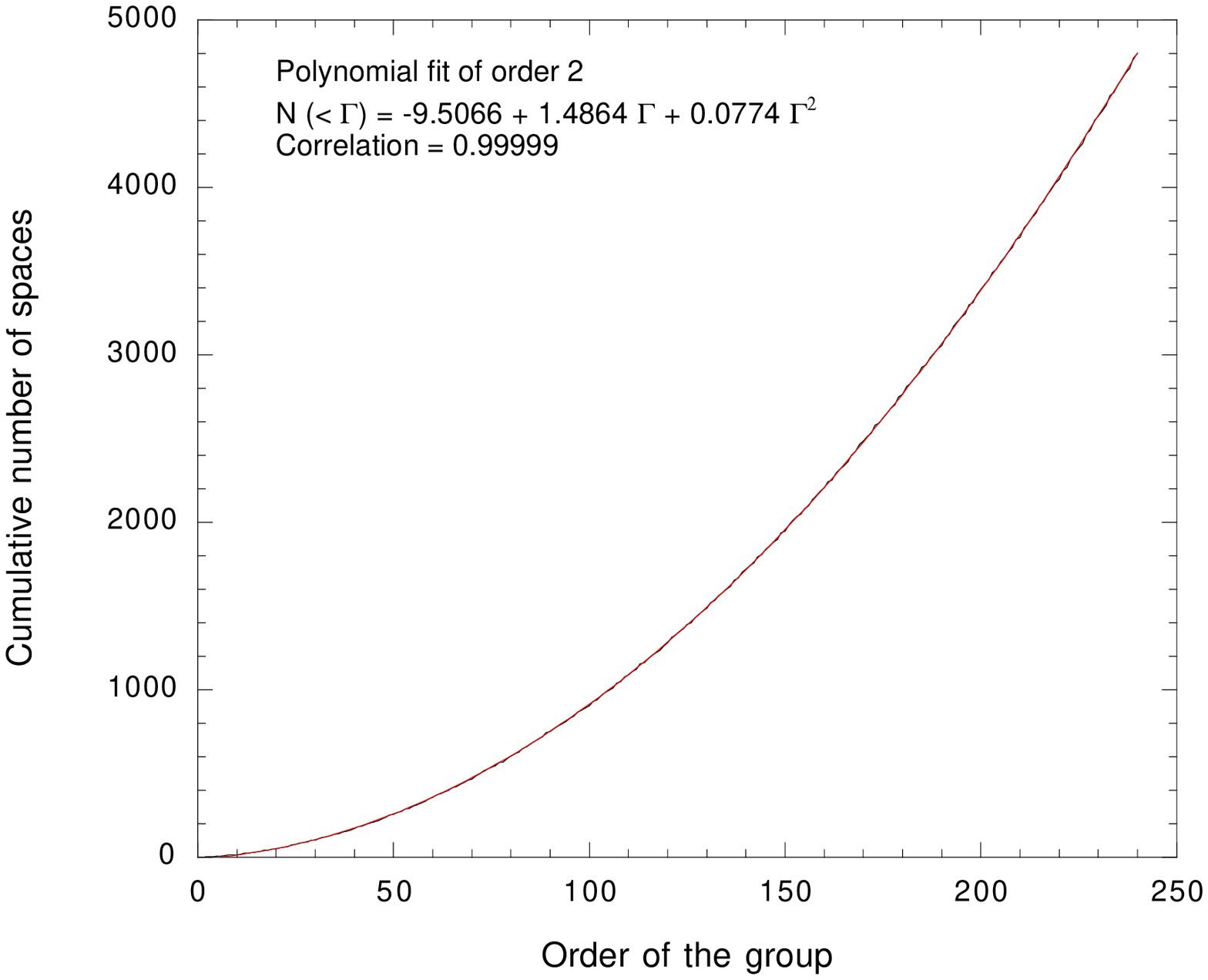, width=7cm}} 
\caption{Left: The number of distinct spherical
3--manifolds of order exactly $|\Gamma|$. The vast majority of these
manifolds are lens spaces for which there are roughly
$|\Gamma|$ choices for the amount of twist in their construction.
So, the number of distinct spherical manifolds of order
exactly $|\Gamma|$ grows linearly with $|\Gamma|$.
Right: The total number of spherical 3--manifolds of order $|\Gamma|$
or less is the sum of an arithmetic series, and therefore grows
quadratically with $|\Gamma|$.} 
\label{number} 
\end{figure}

On Figure~\ref{number}, we show the number of distinct spherical
3--manifolds of a given order. The vast majority of these
manifolds are lens spaces.  This does not necessarily mean that a
spherical universe is ``more likely'' to be a lens space.  It does,
however, reflect the fact that there is a free parameter governing the
amount of twist in the construction of a lens space of order $n$.  The
remaining groups of order $n$ have no free parameters in their
construction.  The difference arises because each lens space is
generated by a single element and is therefore subject to minimal
consistency constraints; the remaining manifolds, which each requires
at least two generators, are subject to stronger consistency
conditions.

In the construction of a lens space of order $n$, there are roughly
$n$ choices for the amount of twist (the exact number of choices
varies with $n$, because the twist parameter must be relatively prime
to $n$), so the number of distinct spherical manifolds of order
exactly $n$ grows linearly with $n$ (see Figure~\ref{number}, left
plot).  Thus the total number of spherical 3--manifolds of order $n$
or less is the sum of an arithmetic series, and therefore grows
quadratically with $n$ (see Figure~\ref{number}, right plot).

\begin{figure}[ht] 
\begin{tabular}{|c|cc|ccc|cc|}
    \hline
    Order & \multicolumn{2}{c|}{Single action} & \multicolumn{3}{c|}{Double action} &
    \multicolumn{2}{c|}{Linked action} \\
    \hline
    1 & $Z_1$ &  &  &  &  &  &  \\
    \hline
    2 & $Z_2$ &  &  &  &  &  &  \\
    \hline
    3 & $Z_3$ &  &  &  &  &  &  \\
    \hline
    4 & $Z_4$ &  &  &  &  &  &  \\
    \hline
    5 & $Z_5$ &  &  &  &  &$Z_5$  &  \\
    \hline
    6 & $Z_6$ &  &  &  &  &  &  \\
    \hline
    7 & $Z_7$ &  &  &  &  &$Z_7$  &  \\
    \hline
    8 & $Z_8$ & $D^*_2$  &  &  &  & $Z_8$ & \\
    \hline
    \ldots &  &  &  &  &  &  &  \\
    \hline
    12 & $Z_{12}$ & $D^*_3$ & $Z_3\times Z_4$ &  &  &  &  \\
    \hline
    \ldots &  &  &  &  &  &  & \\
    \hline
    72 & $Z_{72}$ & $D^*_{18}$  & $Z_8\times Z_9$ & $D^*_2\times Z_9$ &  
       & $Z_{72}$ & $T^*\times Z_9$  \\
       &  &  &  &  &  & $D^*_9\times Z_8$ &  \\
    \hline
    \ldots &  &  &  &  &  &  &   \\
    \hline
    120 & $Z_{120}$ & $D^*_{30}$ & $Z_{40}\times Z_{3}$ & 
          $Z_{24}\times Z_{5}$ & $Z_{8}\times Z_{15}$ & 
          $Z_{120}$ & $D^*_{3}\times Z_{40}$ \\
        & $I^*$ & & $D^*_{10}\times Z_{3}$ & $D^*_{6}\times Z_{5}$ & $D^*_{2}\times Z_{15}$
	    & $D^*_{5}\times Z_{24}$ & $D^*_{15}\times Z_{8}$  \\
        &  &  & $T^*\times Z_{5}$ &  &  &  &  \\
    \hline
    \ldots &  &  &  &  &  &  &  \\
    \hline
    216 & $Z_{216}$ & $D^*_{54}$ & $Z_{8}\times Z_{27}$ & 
        $D^*_{2}\times Z_{27}$  &  & $Z_{216}$ & $T^*\times Z_{27}$ \\
        &  &  &  &  &  & $D^*_{27}\times Z_{8}$ &  \\
    \hline
    \ldots &  &  &  &  &  &  &  \\
    \hline
    240 & $Z_{240}$ &  $D^*_{60}$ & $Z_{80}\times Z_{3}$   & $Z_{48}\times Z_{5}$ &
    $Z_{16}\times Z_{15}$ & $D^*_{20}\times Z_{3}$ & $Z_{240}$ \\

	&  &  & $D^*_{3}\times Z_{80}$ & $D^*_{12}\times Z_{5}$ & $D^*_{4}\times Z_{15}$ 
	& $D^*_{5}\times Z_{48}$ &  $D^*_{15}\times Z_{16}$ \\

	 &  &  & $O^*\times Z_{5}$ &  &  &  &  \\
	\hline
\end{tabular}
\caption{Classification of groups generating single, double and linked
action spherical 3--manifolds.  The first column gives the order
$|\Gamma|$ of the holonomy group, i.e. along each row the volume of
space is $2 \pi^2/|\Gamma|$.  The simply connected 3--sphere is
$S^{3}/Z_{1}$, and the projective space $P^3$ is $S^{3}/Z_{2}$.  Since
some cyclic groups $Z_n$ have different realizations as single action,
double action, and linked action manifolds they may appear more than
once on a given line.  For example, the lens spaces $L(5,1)$ and
$L(5,2)$ are nonhomeomorphic manifolds, even though their holonomy
groups are both isomorphic to $Z_5$.  The double action groups $Z_{m}
\times Z_{n} \simeq Z_{mn}$ all yield lens spaces, as do the single
action groups $Z_{mn}$.  For instance $L(12,1)$ is the single action
manifold generated by $Z_{12}$ while $L(12,5)$ is the double action
manifold generated by $Z_{3} \times Z_{4} \simeq Z_{12}$.  For
noncyclic groups, the realization as a holonomy group is unique, and
thus the resulting manifold is unique, named after its holonomy
group.}
\label{classif}  
\end{figure} 

\begin{figure}[ht]
\begin{center}
\begin{tabular}{|c|c|c|c|}
	\hline
	Order & Single action & Double action & Linked action  \\
	\hline
	2 & $L(2,1)$ &  &   \\
	\hline
	3 & $L(3,1)$ &  &   \\
	\hline
	4 & $L(4,1)$ &  &   \\
	\hline
	5 & $L(5,1)$ &  & $L(5,2)$  \\
	\hline
	6 & $L(6,1)$ &  &   \\
	\hline
	7 & $L(7,1)$ &  & $L(7,2)$  \\
	\hline
	8 & $L(8,1)$ &  & $L(8,3)$  \\
	\hline
	\ldots &  &  &   \\
	\hline
	12 & $L(12,1)$ & $L(12,5)$ &   \\
	\hline
	\ldots &  &  &   \\
	\hline
	72 & $L(72,1)$ & $L(72,17)$ & $L(72,5)$  \\
	  &  &  & + 5 more  \\
	\hline
	\ldots &  &  &   \\
	\hline
	120 & $L(120,1)$ & $L(120,31)$ & $L(120,7)$  \\
	    &            & $L(120,41)$ & + 7 more  \\
	    &            & $L(120,49)$ &   \\
	\hline
	\ldots &  &  &   \\
	\hline
	216 & $L(216,1)$ & $L(216,55)$ & $L(216,5)$  \\
	  &  &  & + 17 more  \\
	\hline
	\ldots &  &  &   \\
	\hline
	240 & $L(240,1)$ & $L(240,31)$ & $L(240,7)$  \\
	    &            & $L(240,41)$ & + 15 more  \\
	    &            & $L(240,49)$ &   \\
	\hline
\end{tabular}
\end{center}
\caption{Classification of lens spaces. In this chart each lens space 
occurs only once in the first valid column (e.g. if the lens space is 
a single action manifold we ignore the trivial expression of it as a 
double or linked action manifold). This also takes into account the 
various equivalences so, for example $L(7,2)$ appears while $L(7,3)$ does 
not because $L(7,3) = L(7,2)$.}
\label{classif2}  
\end{figure} 

\section{Crystallographic Simulations}
\label{IV}  

As we emphasized, the PSH method applies as long as the holonomy 
group has at least one Clifford translation. Thus we use the PSH
method for most of the spherical manifolds but we will 
need the CCP method in certain exceptional cases. 
 
Section~\ref{IVC} determines the radius $\chi_{\rm max}$ of the
observable portion of the covering 3--sphere in units of the curvature
radius, as a function of the cosmological parameters $\Omega_m$ and
$\Omega_{\Lambda}$ and a redshift cutoff $z_c$.  For plausible values
of $\chi_{\rm max}$, Section~\ref{expect} determines which topologies
are likely to be detectable.  Section~\ref{PSHexpectations} explains
the expected form of the Pair Separation Histogram first in the
3--sphere, and then predicts the location and height of the spikes in
a multiply connected spherical universe.  Computer simulations
(\S~\ref{IVA}) confirm these predictions.  In \S~\ref{IVB}, we briefly
recall the applicability of the CCP method.

\subsection{Observational prospects}\label{IVC}  

\begin{figure}[ht] 
\centerline{\psfig{file=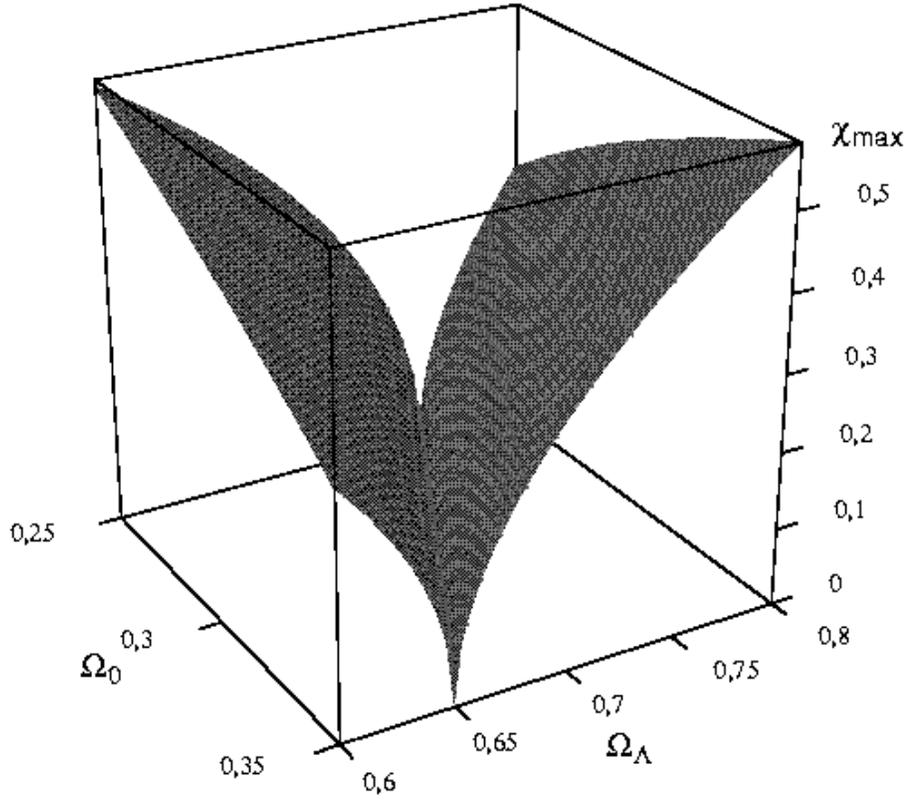, width=13cm}}
\caption{The maximum radial distance $\chi_{\rm max}$ accessible in
a catalog of depth $z_c=5$ as a function of the cosmological
parameters.} 
\label{order} 
\end{figure}

Observations indicate that $\left|\Omega_{m_0}
+\Omega_{\Lambda_0}-1\right|$ is at most about $0.1$, and this value
fixes the physical curvature scale $R_{C_0}^\mathrm{phys}$ from
(\ref{rco}).  To detect topology one requires that the size of the
manifold be smaller than the diameter of the observable universe in at
least one direction.  Equation (\ref{chi2}) gives the distance $\chi$
in radians from the observer to a source at redshift $z$. 
Figure~\ref{order} uses equation (\ref{chi2}) to plot the maximal
radial distance $\chi_{\rm max}$ that is accessible in a catalog of
sources extending to redshift $z_c$, as a function of the cosmological
parameters $\Omega_{m_0}$ and $\Omega_{\Lambda_0}$.  This maximal
radial distance $\chi_{\rm max}$ may also be used to compute the
volume of the observable universe ${\rm vol}(\chi_{\rm max}) = \pi(2
\chi_{\rm max} - \sin 2 \chi_{\rm max}) \simeq \frac{4}{3} \pi
\chi_{\rm max}^3$ in curvature radius units and compare it to the
total volume $2 \pi^2$ of the 3--sphere.

\begin{figure}[ht] 
\centerline{\epsfig{file=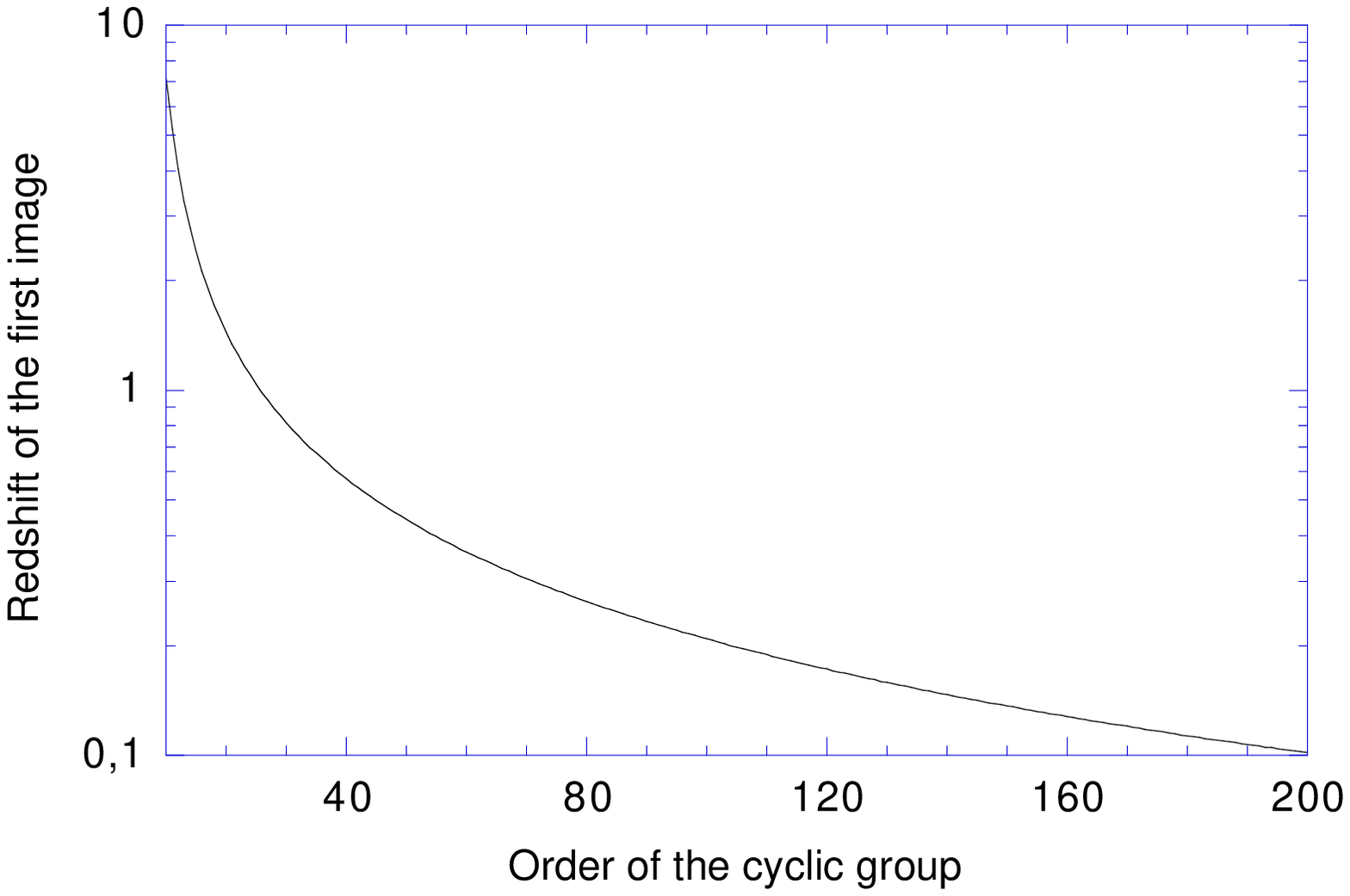, width=8cm}} 
\caption{The redshift of the first topological image as a function of
the order of the cyclic group $Z_{n}$ when $\Omega_0=0.35$ and
$\Omega_{\Lambda_0}=0.75$.} 
\label{zfirst} 
\end{figure}

In practice one requires that shortest distance between topological
images be less than the effective cutoff radius $\chi_{\rm max}$.
For example, a cyclic group $Z_n$ will satisfy this criterion if and
only if
\begin{equation}
\frac{2\pi}{n} < \chi_{\rm max}.
\end{equation}
Figure~\ref{zfirst} shows the implications of this equation.
In principle one could detect topology even if the shortest
translations were almost as large as the diameter of the
observable universe, i.e. $2 \chi_{\rm max}$, but when using
statistical methods the signal would become too weak.

\subsection{Geometrical expectations}\label{expect}

In this section we determine which topologies are likely
to be detectable for plausible values of $\chi_{\rm max}$ (see also 
\cite{gom01} for detectability of lens spaces).

In a single action manifold every holonomy is a Clifford
transformation, so in principle the PSH detects the entire holonomy
group.  In practice we see only a small portion of the covering
3--sphere.  For example, with $\Omega_{m_0} = 0.35$,
$\Omega_{\Lambda_0} = 0.75$, and a redshift cutoff of $z = 3$, we see
a ball of radius $\chi_{\rm max} \simeq{1}/{2}$ (see
figure~\ref{order}).  With this horizon size the binary tetrahedral,
binary octahedral, and binary icosahedral groups would be hard to
detect because their shortest translations distances are ${2\pi}/{6}$,
${2\pi}/{8}$, and ${2\pi}/{10}$, respectively.  If, however, we extend
the redshift cutoff to $z=1000$, then the horizon radius expands to
$\chi\simeq 1$, putting the binary tetrahedral, binary octahedral, and
binary icosahedral groups within the range of CMB methods.

The cyclic groups $Z_n$ and the binary dihedral groups $D^\ast_m$ are
much more amenable to detection, because the shortest translation
distances, ${2\pi}/{n}$ and ${2\pi}/{2 m}$ respectively, can be
arbitrarily small for sufficiently large $n$ and $m$.  For the example
given above with $\chi_{\rm max}\simeq{1}/{2}$, a cyclic group $Z_n$
will be detectable if $n>{2\pi}/({1}/{2}) \simeq 12$, and similarly a
binary dihedral group will be detectable if $2m > 12$.  On the other
hand, the lack of obvious periodicity on a scale of 1/20 the horizon
radius\footnote{Note that this upper bound is very approximate and
that a detailed study will be required.} implies that $n$ and $2m$ may
not exceed ${2\pi}/({1}/{20}) \simeq 120$.  In conclusion, the orders
of the cyclic groups $Z_n$ and the binary dihedral groups $D^\ast_m$
that are likely to be detectable must lie in the ranges
$$12\lsim n\lsim 120\quad\mathrm{and}\quad 12\lsim 2m\lsim 120.$$
The holonomy group of a double action manifold (recall \S~\ref{da})
consists of the products $r l$, as $r$ ranges over a group $R$ of
right--handed Clifford translations and $l$ ranges over a group $L$ of
left--handed Clifford translations.  Because only pure Clifford
translations generate spikes, the PSH detects a product $r l$ if and
only if $r = \pm{{\bf 1}}$ or $l = \pm{{\bf 1}}$.  Thus the set of
spikes in the PSH for the double action manifold defined by $R$ and
$L$ will be the union of the spikes for the two single action
manifolds defined by $R$ and $L$ alone, or possibly
$R \times \lbrace \pm{{\bf 1}} \rbrace$ or
$L \times \lbrace \pm{{\bf 1}} \rbrace$.  For
example, if $R = D^*_2$ and $L = Z_3$, then we get the union of the
spikes for $D^*_2$ with the spikes for $Z_3\times \lbrace \pm{{\bf 1}}
\rbrace = Z_6$ (see figure~\ref{psh1}).  The conditions for
observability thus reduce to the conditions described for a single
action manifold.  That is, a double action manifold is most easily
detectable if one of its factors (possibly extended by $-{\bf 1}$)
is a cyclic group $Z_n$ or a binary dihedral group $D^\ast_m$, with 
$$12 \lsim n \lsim 120\quad\mathrm{and}\quad12\lsim 2m \lsim 120.$$
Happily the shortest translations are all Clifford translations.

In a linked action manifold (see \S~\ref{la}) not every element of $R$
occurs with every element of $L$.  Let $R'$ be the subgroup consisting
of those elements $r \in R$ that occur with $\pm{{\bf 1}}$ of
$L$, and let $L'$ be the subgroup consisting of those elements $l \in
L$ that occur with $\pm{\bf 1}$ of $R$.  With this notation,
the Clifford translations are the union of $R'$ and $L'$, possibly
extended by $-{\bf 1}$.

Linked action manifolds are the most difficult to detect
observationally, because the groups $R'$ and $L'$ may be much smaller
than $R$ and $L$.  Indeed for some linked action manifolds the groups
$R'$ and $L'$ are trivial.  For example, the lens space $L(5,2)$ is
obtained as follows.  Let $R = Z_5$ and $L = Z_5$, and let the
preferred generator $r$ of $R$ be a right-handed Clifford translation
through an angle $2\pi/5$ while the preferred generator $l$ of $L$ is
a left-handed Clifford translation through an angle $4\pi/5$.  Link
the generator $r$ to the generator $l$, and link their powers
accordingly.  The identity element of $R$ (resp.  $L$) occurs only
with the identity element of $L$ (resp.  $R$), so both $R'$ and $L'$
are trivial.  In this case the PSH contains no spikes and we must use
the CCP method or CMB methods instead.\\

We emphasize that the detectability of a given topology depends not on
the order of the group, but on the order of the cyclic factor(s).
Given that we can see to a distance of about $2\pi/12$ in comoving
coordinates (see discussion above), the lens space $L(17,1)$, a single
action manifold of order 17, would be detectable.  However, the lens
space $L(99,10)$, a double action manifold of order 99 generated by
$Z_9\times Z_{11}$, would probably not be detectable, because its
shortest Clifford translations have distance $2\pi/11$.  The only
way we might detect $L(99,10)$ would be if the Milky Way, just by
chance, happened to lie on a screw axis, where the flow lines of the
right-- and left--handed factors coincide.  On the other hand, a double
action manifold like $S^3/(Z_{97}\times Z_{98})$ probably is detectable.  It
has Clifford translations of order 97 and 98, so crystallographic
methods should in principle work, and it does not violate the observed
lack of periodicity at $1/20$ the horizon scale because its screw axis
(along which the minimum translation distance is $2\pi/(97*98) =
2\pi/9506$) probably comes nowhere near the Milky Way.\\

Putting the results all together and using the complete classification
of spherical spaces, we estimate that a few thousands of potentially
observable topologies are consistent with current data.  This number
is reasonably low in view of search strategies for the detection of
the shape of space through crystallogaphic or CMB methods.  We note
that similar lower bounds were recently obtained in~\cite{gom01} in
the particular case of lens spaces, and this analysis does not give
upper bounds and is less general than the one presented here.

\subsection{PSH spectra}
\label{PSHexpectations}  

Let us now discuss the form of ideal PSH spectra. We start by
considering the expected PSH spectrum in a simply--connected spherical
space. As shown in~\cite{bernui}, it is obtained by computing the
probablility ${\cal P}(\chi_a,\chi_l)\dd\chi_l$ that two points in a
ball of radius $\chi_a$ are separated by a distance between $\chi_l$
and $\chi_l+\dd\chi_l$
\begin{eqnarray}
{\cal P}(\chi_a,\chi_l) &=& \frac{8\sin^2\chi_l}{[2\chi_a-\sin 2\chi_a]^2}
                            \{[2\chi_a-\sin 2\chi_a-\pi]
                            + \nonumber\\  
                        && \Theta(2\pi-2\chi_a-\chi_l)\times 
                           [\:\sin\! 2\chi_a+\pi-\chi_a-
                           \chi_l/2\nonumber\\
                        &&-\cos\!\chi_a\: \sec(\chi_l/2)
                        \:\sin(\chi_a-\chi_l/2)]\},
\end{eqnarray}
valid for all $\chi_a\in [0,\pi]$ and $\chi_l\in [0,{\rm
min}(2\chi_a,\pi)]$ and in which $\Theta$ stands for the Heavyside
function. 

As seen in the sample PSH spectra of figure~\ref{psh1},
the PSH of the simply--connected space $S^3$ provides the
background contribution of the Poisson distribution
over which the spikes of the multi--connected space
will appear.

Each Clifford translation $g$ of the holonomy group will generate a
spike in the PSH spectrum. A Clifford translation takes the
form $M(\theta,\theta)$ or $M(\theta, - \theta)$ (see \S~\ref{generalities})
and so its corresonding spike occurs at $\chi=\theta$. The number of group
elements sharing the same value of $|\theta|$, with $\theta$ normalized
to the inverval $[-\pi,\pi]$, is the multiplicity ${\rm mult}(\theta)$
of the translation distance $\theta$.

To compute the amplitude of the spike, we first consider the case
that there is only one source in the fundamental polyhedron.
In that case, the number of images in the covering 3--sphere equals the order
of the holonomy group $|\Gamma|$. Each single image sees ${\rm mult}(\theta)$
neighbours lying at distance $\theta$ from it, so that this distance
appears $|\Gamma|\,{\rm mult}(\theta)/2$ times (the division by two
compensates for the fact that we counted the distance from image A to
image B as well as the distance from image B to image A). We conclude
that if we have $N$ sources in the fundamental polyhedron, the
amplitude of the spike located at $\chi=\theta$ is
$$
{\rm amplitude}(\theta)=\frac{1}{2}\,N\,|\Gamma|\,{\rm mult}(\theta).
$$
As an example, let us consider the case of the cyclic group $Z_6$
consisting of six Clifford translations through distances
$(-2\pi/3,-\pi/3,0,\pi/3,2\pi/3,\pi)$.
These Clifford translations yield spikes of multiplicity 1 at
$\chi = 0$ and $\chi = \pi$, and spikes of multiplicity 2 at
$\chi = \pi/3$ and $\chi = 2\pi/3$, as shown in table~\ref{table2}.
Applying the above formula to the spike at $\chi = \pi/3$,
in the case of $N = 300$ distinct sources in the fundamental domain,
we expect the spike to reach a height
$\frac{1}{2}\,N\,|\Gamma|\,{\rm mult}(\theta)
= \frac{1}{2}(100)(6)(2) = 1800$ above the background distribution,
in agreement with computer simulations (see figure~\ref{pshn}).

In tables \ref{table1} to \ref{table4}, we give the translation
distances $|\theta|$ and the multiplicities ${\rm mult}(\theta)$ for
the binary dihedral group $D^*_2$, the cyclic group $Z_6$, the product
$D_2^*\times Z_{3}$, and the binary icosahedral group $I^*$,
respectively.  They correspond to the PSH spectra shown in
figures~\ref{psh1} and \ref{psh3}.  Note that the spectrum for
$D_2^*\times Z_{3}$ (table \ref{table3}) is obtained from those of the
binary dihedral group $D^*_2$ (table \ref{table1}) and
$Z_3\times\lbrace\pm{{\bf 1}}\rbrace=Z_6$ (table \ref{table2}).

\begin{table}
\begin{center}
\begin{tabular}{|c|ccc|}
\hline
$\chi/\pi$ & 0 & 1/2 & 1 \\
\hline
multiplicity & 1 & 6 & 1 \\
\hline
\end{tabular}
\end{center}
\caption{Position and the multiplicity of the spikes for the 
binary dihedral group $D^*_2$. Compare with
figure~\ref{psh1}.}
\label{table1}
\end{table}

\begin{table}
\begin{center}
\begin{tabular}{|c|cccc|}
\hline
$\chi/\pi$ &0& 1/3 & 2/3 & 1\\
\hline
multiplicity &1& 2& 2&1\\
\hline
\end{tabular}
\end{center}
\caption{Position and the multiplicity of the spikes for the binary
cyclic group $Z_6$.  Compare with figure~\ref{psh1}.}
\label{table2}
\end{table}

\begin{table}
\begin{center}
\begin{tabular}{|c|ccccc|}
\hline
$\chi/\pi$ &0& 1/3 & 1/2 & 2/3 & 1\\
\hline
multiplicity & 1 & 2 & 6 & 2 & 1 \\
\hline
\end{tabular}
\end{center}
\caption{Position and the multiplicity of the spikes for $D_2^*\times
Z_{3}$.  As expected, we find that it can be obtained from the data
for the binary dihedral group $D^*_2$ (table \ref{table1}) and
$Z_3\times\lbrace\pm{{\bf 1}}\rbrace=Z_6$ (table \ref{table2}).  Compare with
figure~\ref{psh1}.}
\label{table3}
\end{table}

\begin{table}
\begin{center}
\begin{tabular}{|c|ccccccccc|}
\hline
$\chi/\pi$ &0& 1/5 & 1/3 & 2/5 & 1/2 & 3/5 & 2/3 & 4/5 & 1 \\
\hline
multiplicity & 1 & 12 & 20 & 12 & 30 & 12 & 20 & 12 & 1\\
\hline
\end{tabular}
\end{center}
\caption{Position and the multiplicity of the spikes for the binary
icosahedral group $I^*$.  Compare with figure~\ref{psh2}.}
\label{table4}
\end{table}

\subsection{PSH numerical Simulations}\label{IVA} 
 
The previous section presented the theoretical expectations for PSH in
multiply connected spherical spaces.  Here we carry out numerical
simulations confirming those expectations.  Later in this section we
will also take into account the approximate flatness of the observable
universe, which implies that we are seeing only a small part of the
covering space, and therefore can observe spikes only at small
comoving distances $\chi$.

\begin{figure}[ht] 
\centerline{\epsfig{file=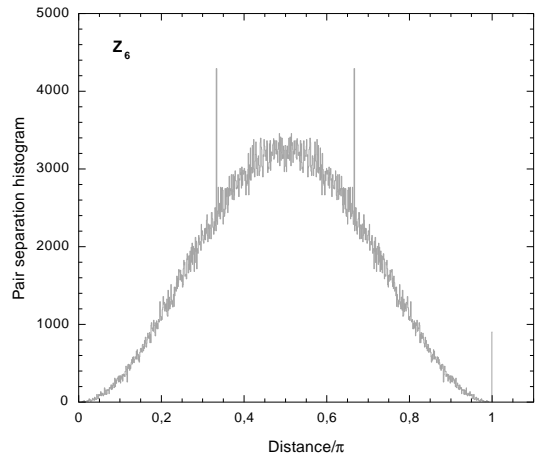, width=7cm}}
\caption{The PSH for the lens space  $L(6,1)$
generated by the cyclic group $Z_{6}$ with $N = 300$ sources 
in the fundamental domain. The height of the peaks at
$\chi=\pi/3$ and $\chi = 2\pi/3$ is of order 4200 while the
background distribution is of order 2400, giving the peaks a height above the
background of about 1800, in agreement with our theoretical analysis.}
\label{pshn} 
\end{figure}

In figure~\ref{pshn} we draw the histogram for the lens space $L(6,1)$
and we check that our geometrical expectations about the positions and
the heights of the spikes are satisfied.
 
\begin{figure}[ht] 
\centerline{\epsfig{file=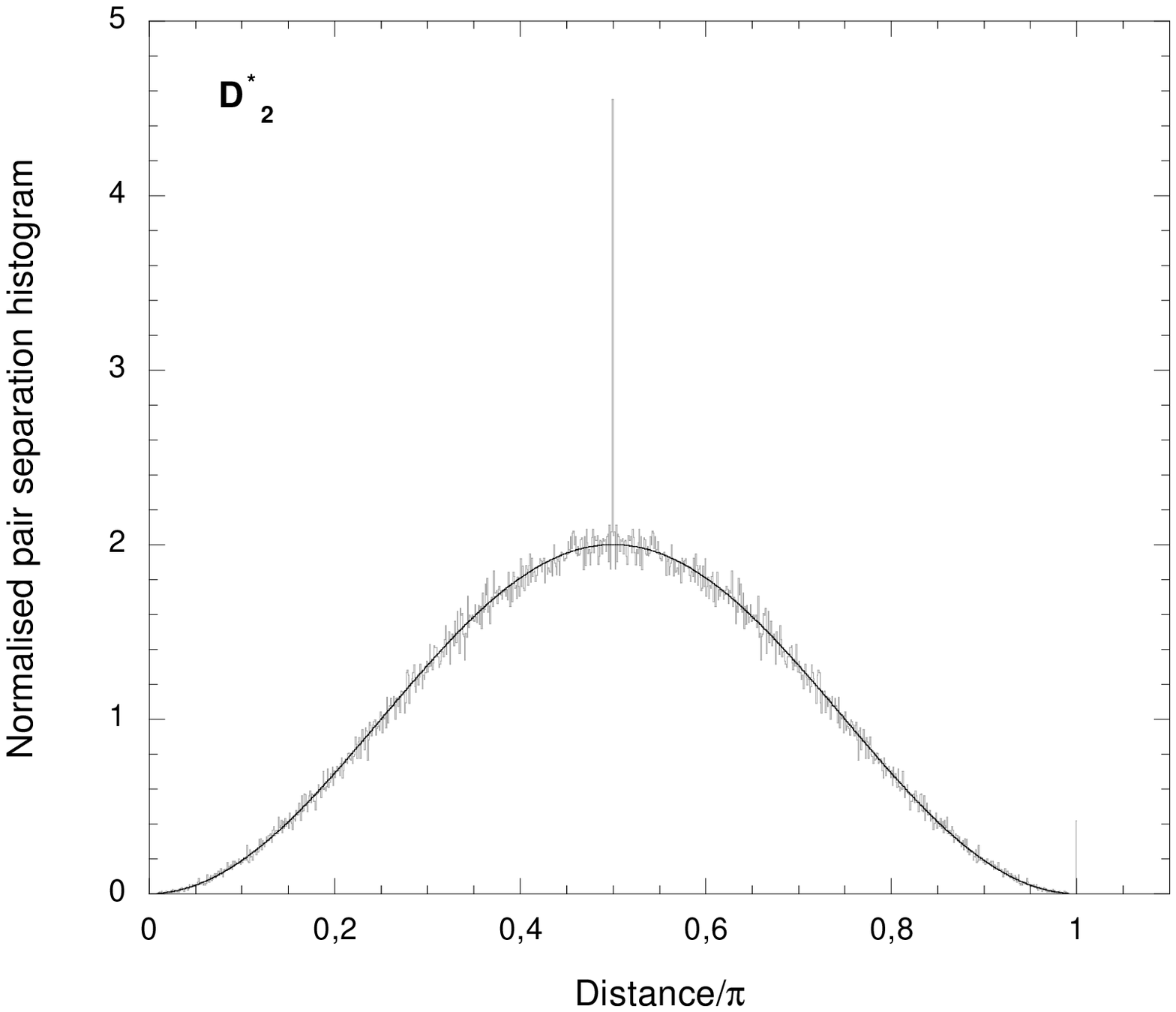, width=7cm}
\epsfig{file=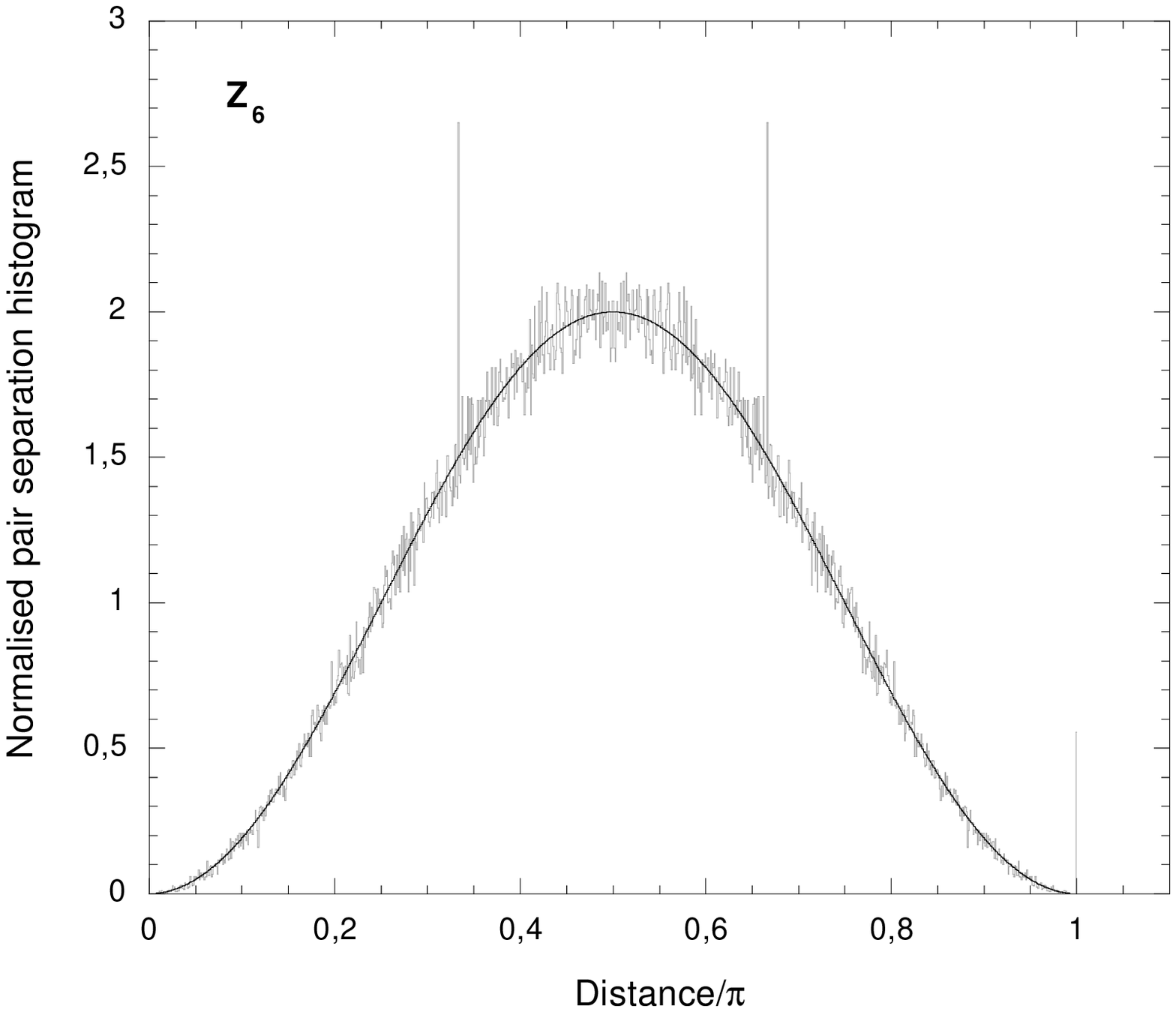, width=7cm}}
\centerline{\epsfig{file=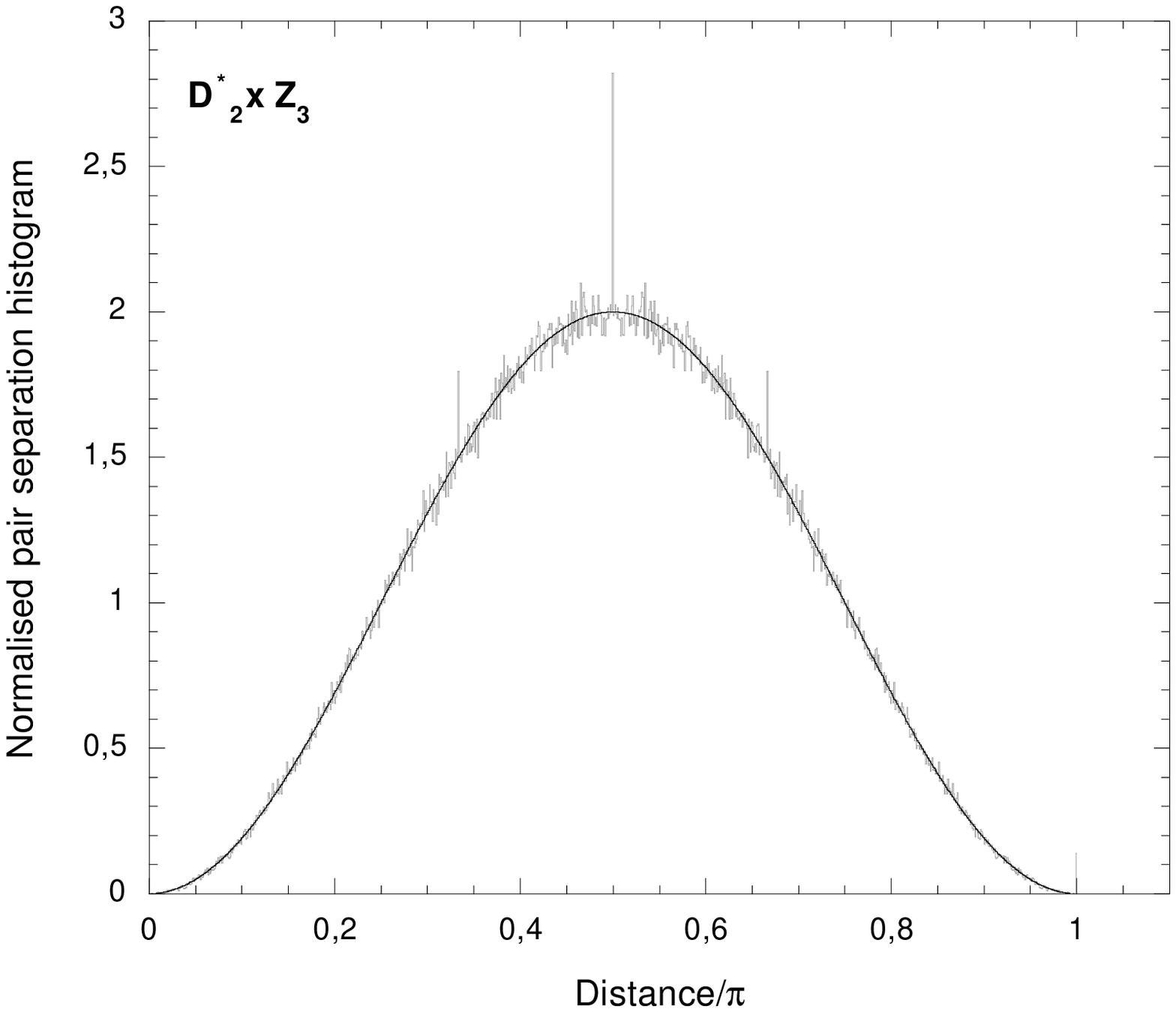, width=7cm}}
\caption{The PSH spectra for (upper left) the binary dihedral group
$D_2^*$, (upper right) the cyclic group $Z_6$ and (bottom)
the group $D_2^*\times Z_3$. One can trace the spikes from each
subgroup and the black line represents the analytic distribution in a
simply--connected universe.}
\label{psh1} 
\end{figure}

\begin{figure}[ht] 
\centerline{\epsfig{file=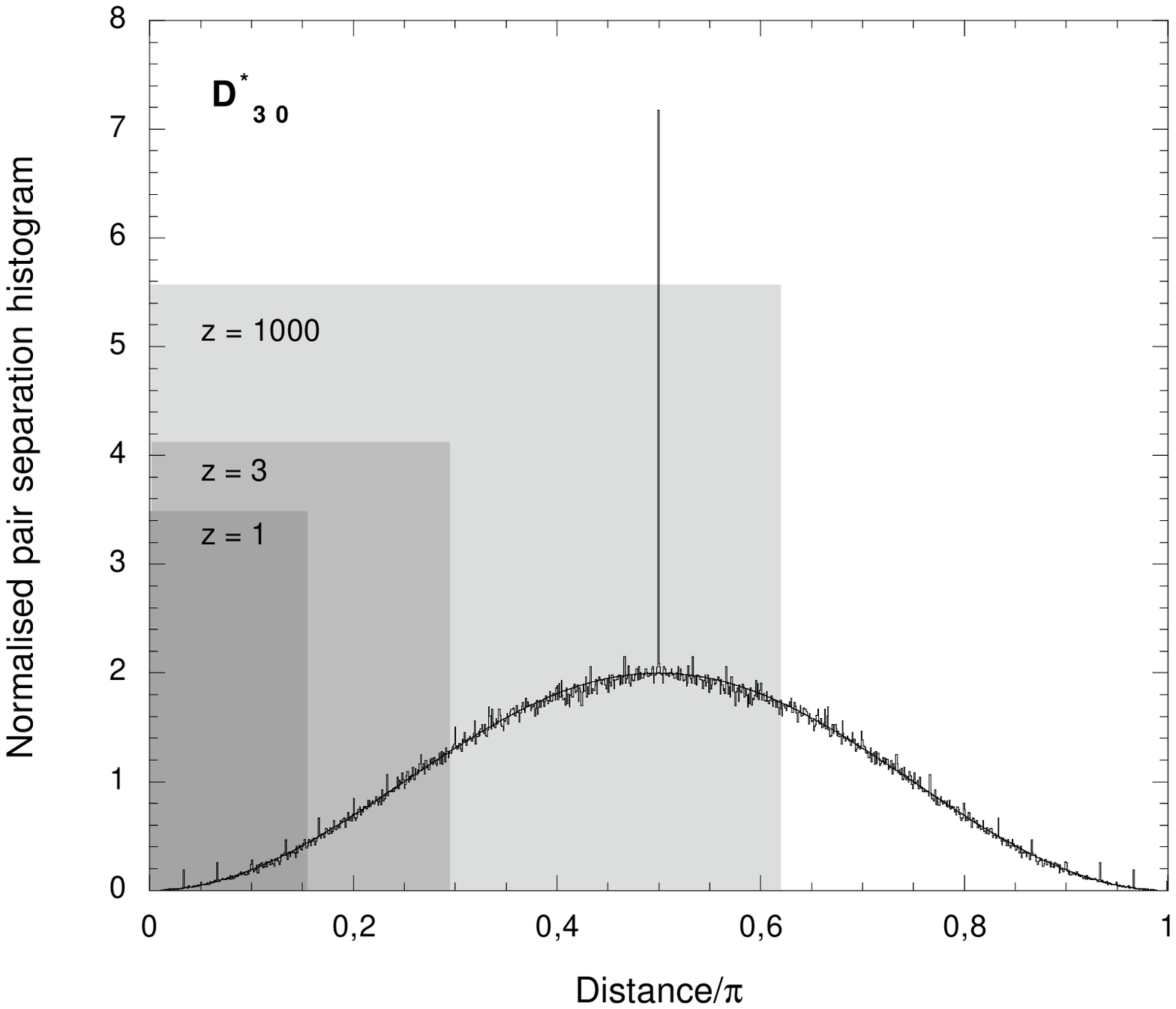, width=7cm}
\epsfig{file=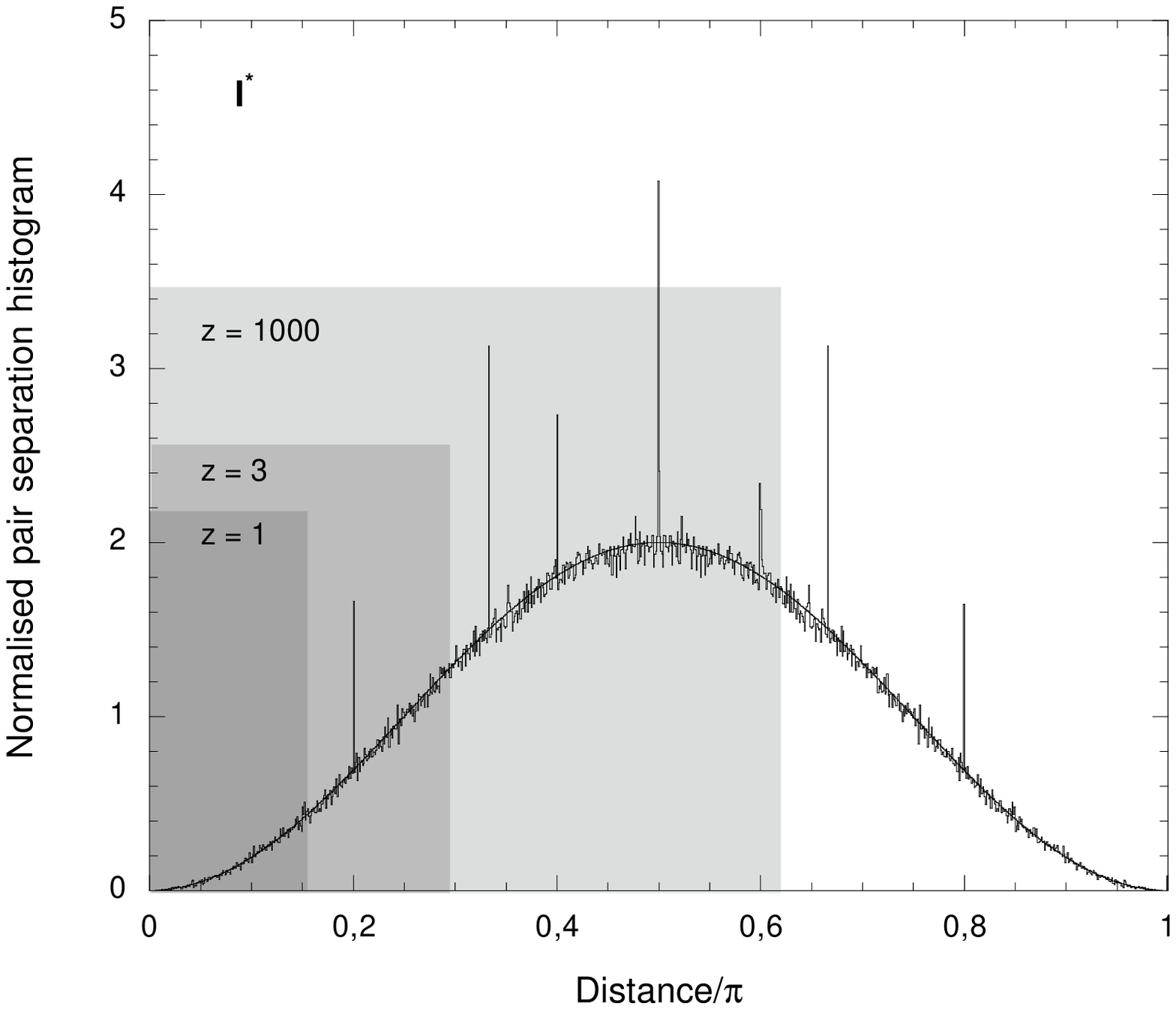, width=7cm}}
\caption{The PSH spectra for two spherical spaces of order 120 but
with no cyclic factor: (left) the binary dihedral group $D_{30}^*$ and
(right) the binary icosahedral group $I^*$}
\label{psh2} 
\end{figure}

\begin{figure}[ht] 
\centerline{\epsfig{file=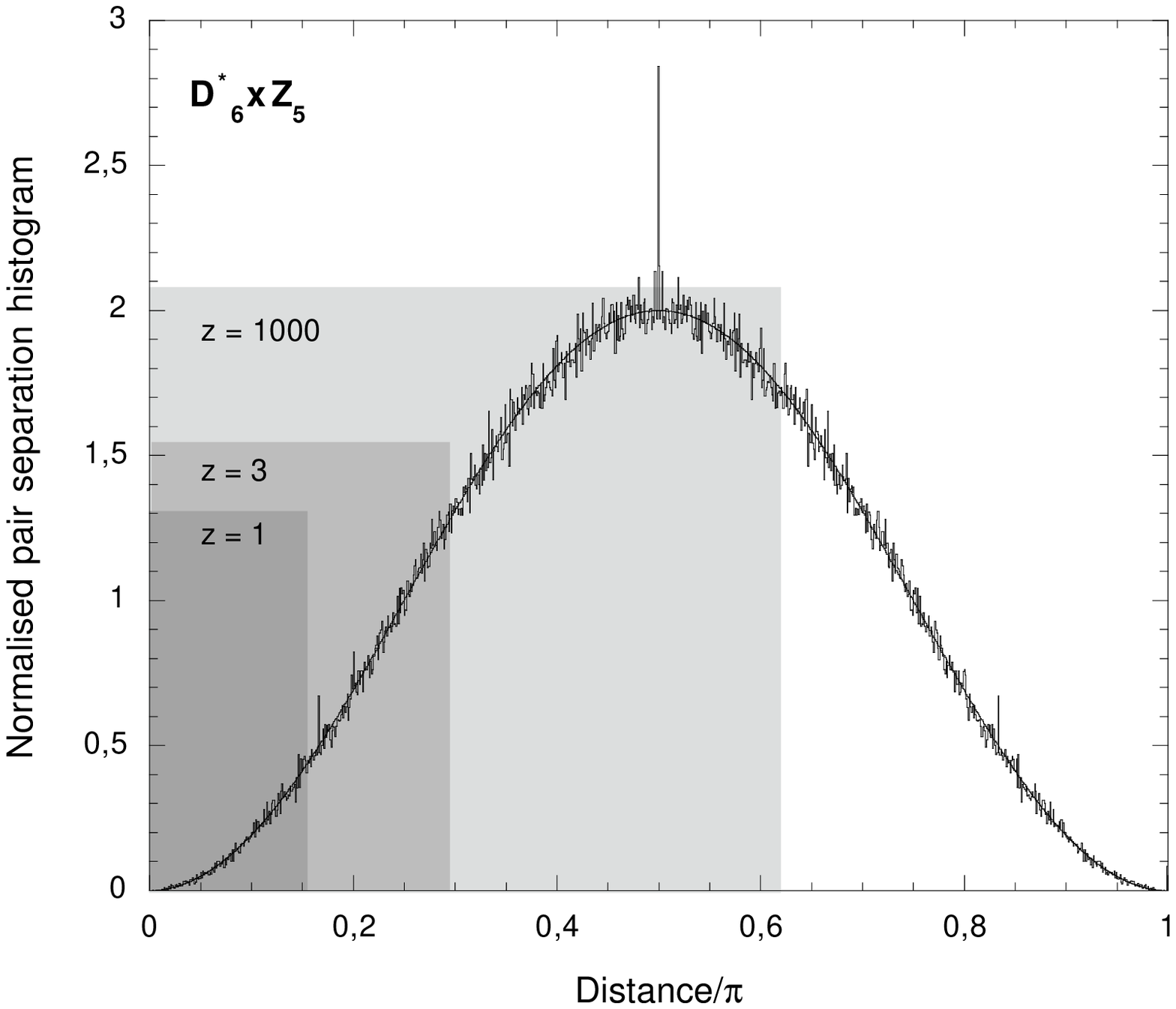, width=7cm}
\epsfig{file=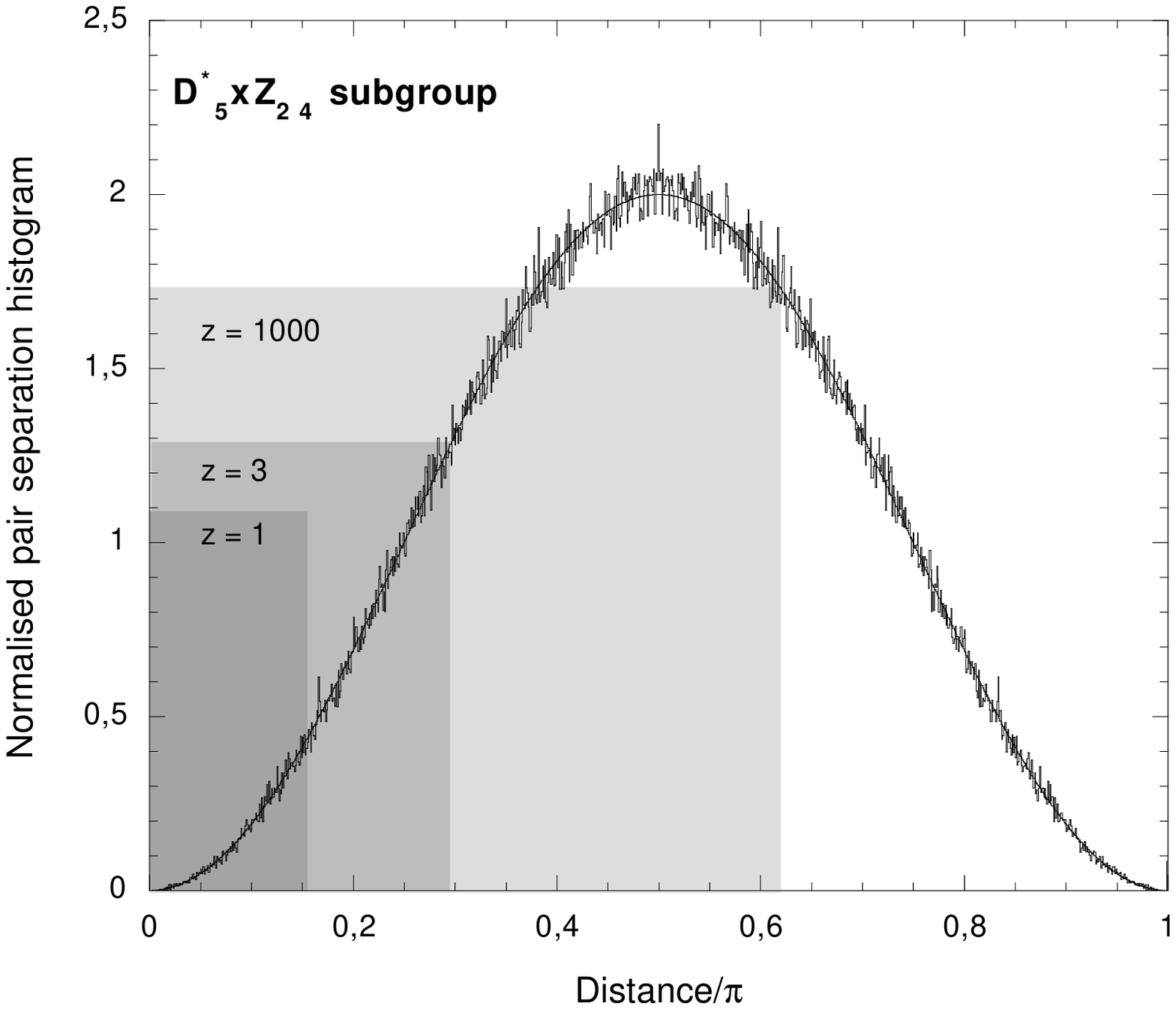, width=7cm}}
\centerline{\epsfig{file=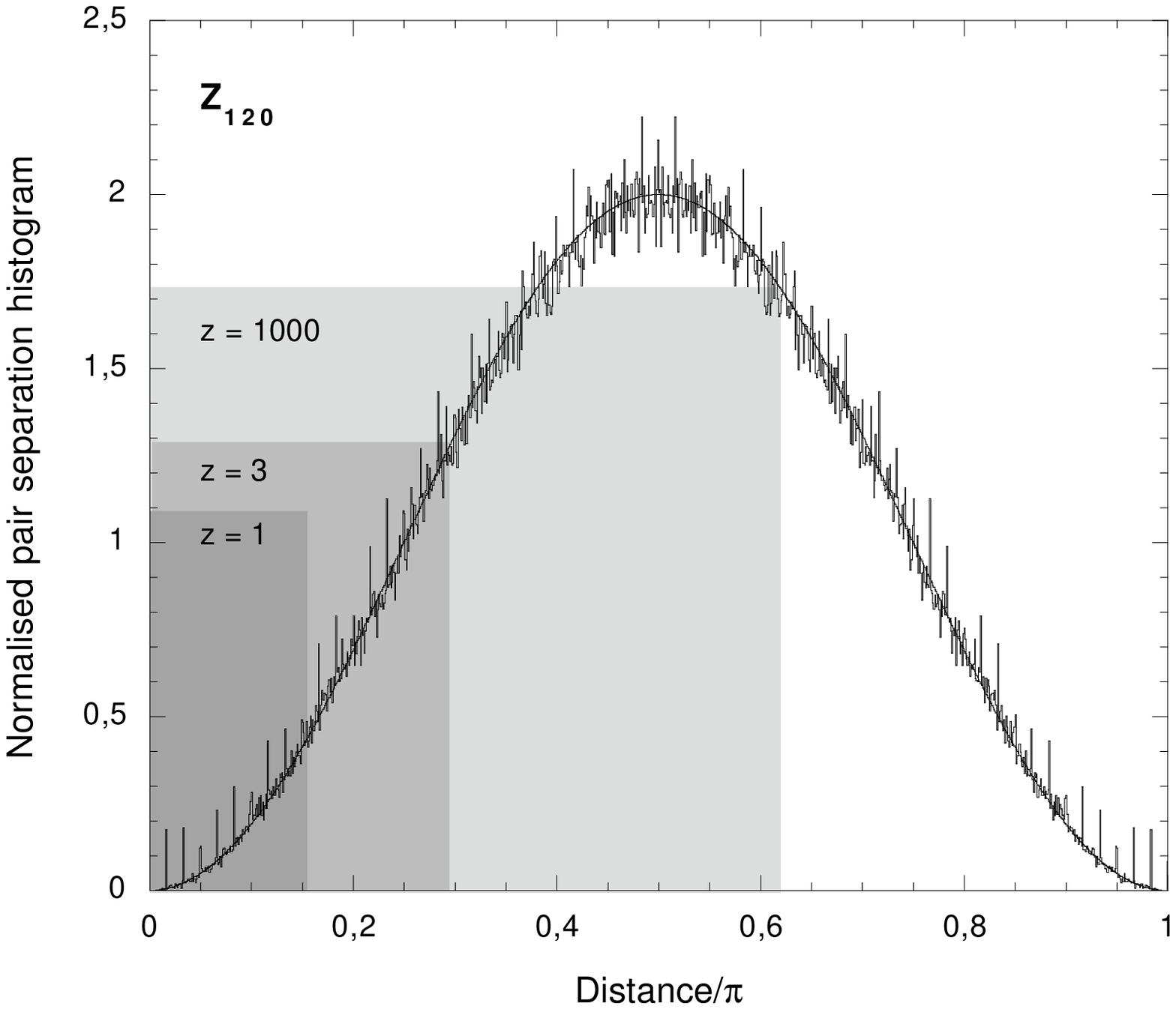, width=7cm}}
\caption{The PSH spectra for three spaces of order 120 with different
cyclic factors: (upper left) $D_6^*\times Z_5^*$, (upper right) $D_5^*\times Z_{12}^*$
 and (bottom) $Z_{120}$. As explained in
the text, the order of the group being fixed, the higher the order of
the cyclic subgroup the larger the number of spikes. However the
amplitude of these spikes is smaller.}
\label{psh3} 
\end{figure}

In figures~\ref{psh1} to \ref{psh3} we present various simulations. 
We have chosen to draw the normalised PSH in the covering space, i.e.
the PSH divided by the number of pairs and the width of the bin, and
to show in grey regions the part of the PSH that is observable if the
redshift cut--off is respectively $z_c=1,3,1000$, which roughly
corresponds to catalogs of galaxies, quasars and the CMB.

We start by showing in figure~\ref{psh1} the ``additivity'' of the
spectrum by considering the group $D_2^*\times Z_{3}$ for which the
positions of the spikes in its PSH can be found from the
PSH of the binary dihedral group $D^*_2$ and of
$Z_3\times\lbrace\pm{{\bf 1}}\rbrace=Z_6$. Indeed, the amplitude of the
spikes is smaller since the same number of pairs has to contribute to
more spikes.

\begin{figure}[ht] 
\centerline{\psfig{file=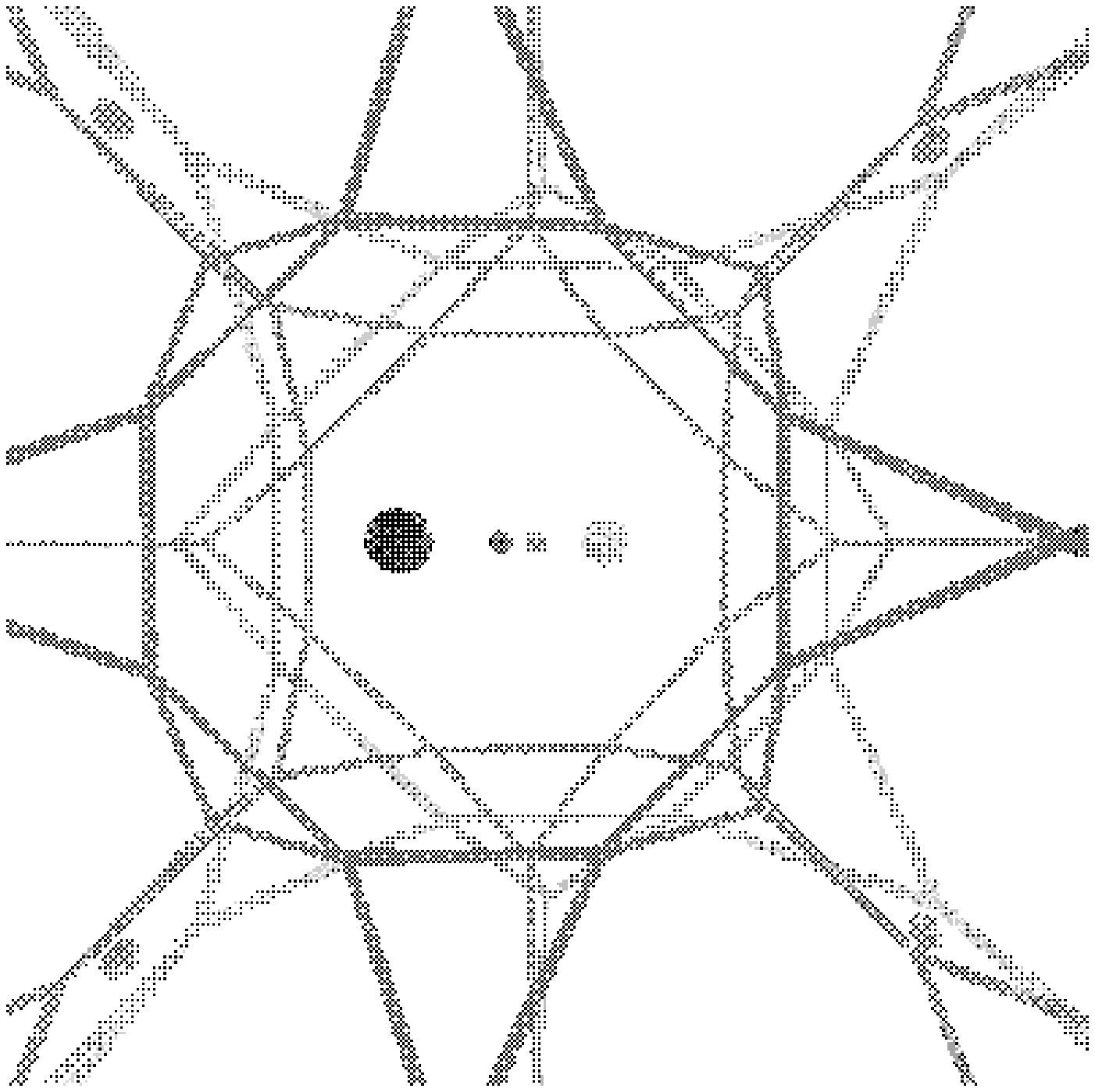, width=6cm}
            \psfig{file=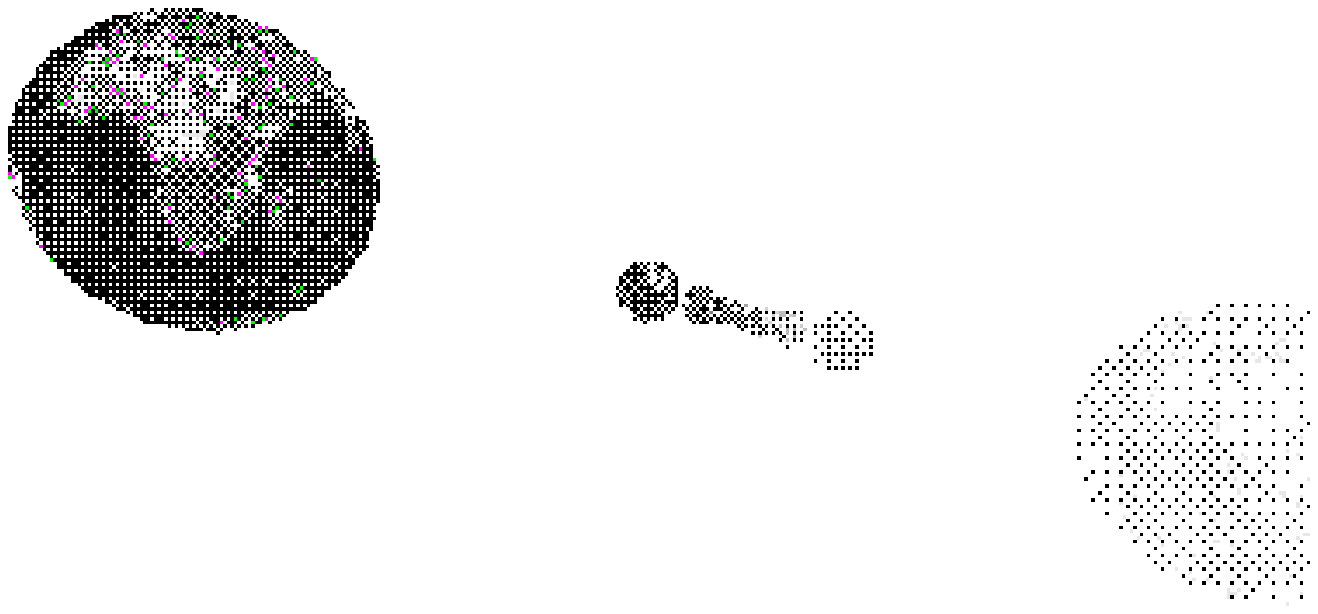, width=6cm}}
\centerline{\psfig{file=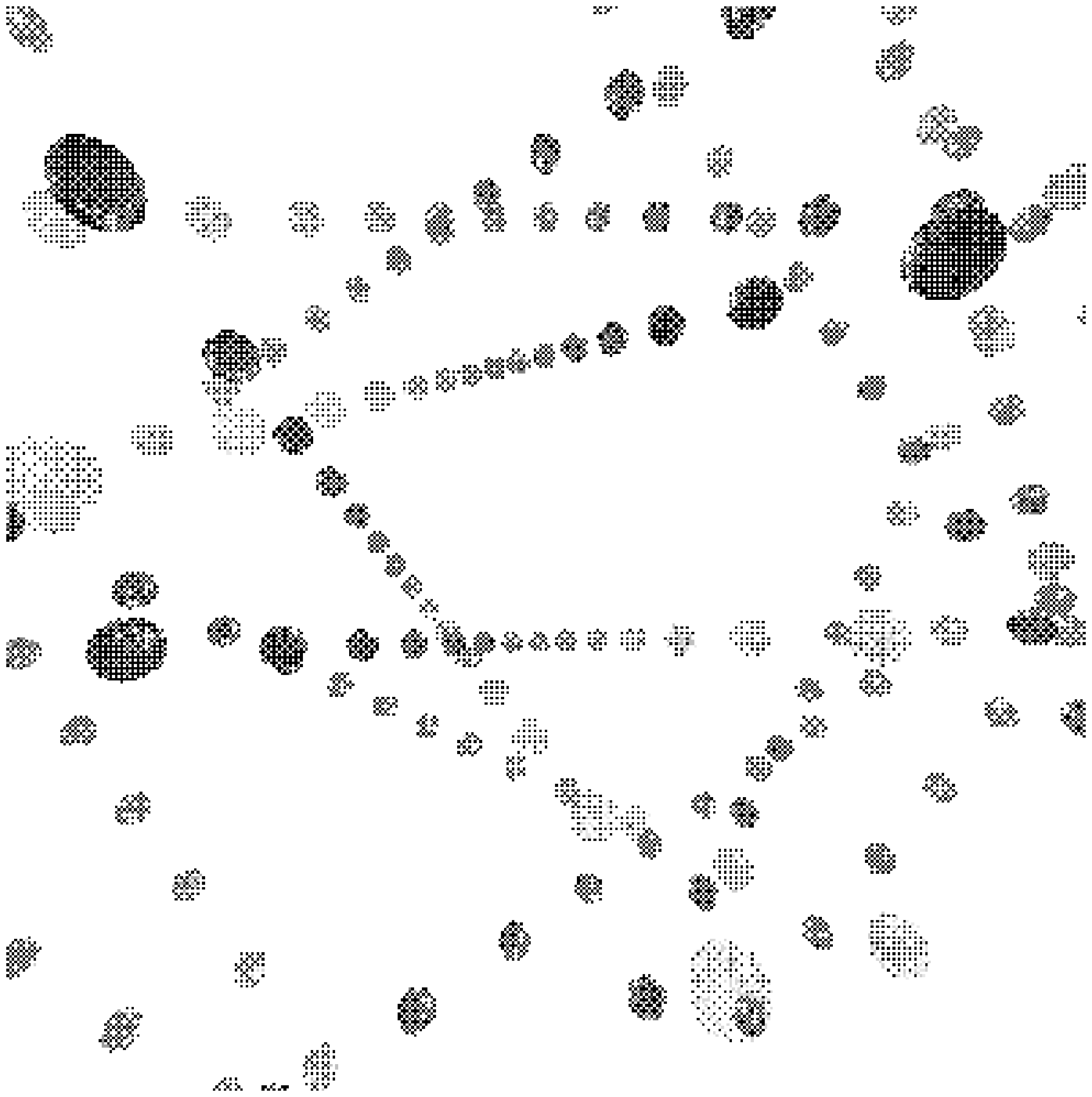, width=6cm}}
\caption{To have an intuitive understanding of which subgroups give
rise to spikes and how, it is fruitful to have a look from the inside.
In the upper left panel, we see the view inside the quotient of the
3--sphere $S^3$ by the binary octahedral group $O^*$.  The fundamental
domain is a truncated cube, 48 copies of which tile $S^3$.  The
tiling's 1--dimensional edges are shown in the figure.  In the upper left
panel, we see the view in $L(17,1)$.  All 17 translates of the Earth
align along a Clifford parallel.  When mixing both groups to get
$O^*\times Z_{17}$ in the bottom panel, we see the simultaneous
effects of the cyclic factor $Z_{17}$, which generates lines of
images, and the binary octahedral factor, which translates one line of
images to another.  This illustrates how important the order of the
cyclic group is, because it alone determines the distance to our
nearest translate.

Note:  More distant images of the Earth are always dimmer than closer
images (because of artificial ``fog''), but the apparent size of an
image decreases only until the images reaches a distance of $\pi/2$,
after which more distant images appear larger because the light from
them follows geodesics that reconverge in $S^3$.  As an image approaches
the antipodal point, at a distance of $\pi$, it fills the whole sky.}
\label{inside} 
\end{figure}

We then show in figures~\ref{psh2} and \ref{psh3} different groups of
the same order, chosen to be 120. The aim here is first to show that
the number of spikes is not directly related to the order of the group
and that the order of the cyclic factor, if any, is more
important. While increasing this order, the number of spikes grows but
their amplitudes diminish. To get a physical understanding of this
effect, we show in figure~\ref{inside} the view for an observer
inside the manifold for the binary octahedral group $O^*$, for the
cyclic group $Z_{17}$, and then for their product.

Finally, we applied our calculations by applying the PSH to real
data, namely a catalog of about 900 Abell and ACO clusters with
published redshifts.  The depth of the catalog is $z_{cut} = 0.26$,
corresponding to $730 h^{-1}$ Mpc in a spherical universe
(see~\cite{lehoucq96} for a more detailed description of this
catalog).  The PSH exhibits no spike; this gives a constraint on the
minimal size of spherical space, which corresponds to a maximum order
of the cyclic factor of the holonomy group of about 200.

\subsection{Robustness of PSH}\label{IVB} 

In \cite{lehoucq00} we addressed the question of the stability of 
crystallograpic methods, i.e. to what extent the topological signal 
is robust for less than perfect data. We listed the 
various sources of observational uncertainties in
catalogs of cosmic objects as:
\begin{enumerate}   
\item[(A)] the errors in the positions of observed objects, namely   
  \begin{enumerate}   
         \item[(A1)] the uncertainty in the determination of the redshifts   
         due to spectroscopic imprecision,   
         \item[(A2)] the uncertainty in the position due to peculiar   
         velocities of objects,    
         \item[(A3)] the uncertainty in the cosmological parameters, which   
         induces an error in the determination of the radial distance,  
         \item[(A4)] the angular   
         displacement due to gravitational lensing by large scale   
         structure,  
\end{enumerate}   
\item[(B)] the incompleteness of the catalog, namely 
\begin{enumerate}   
  \item[(B1)] selection effects,  
  \item[(B2)] the partial coverage of the celestial sphere. 
\end{enumerate}   
\end{enumerate}
 
We showed that the PSH method was robust to observational imprecisions, but
able to detect only topologies whose holonomy groups contain
Clifford translations, and we performed numerical calculations in 
Euclidean topologies.
Fortunately the shortest translations in spherical universes are
typically Clifford translations (even though more distant translations
might not be), so the PSH is well suited to detecting spherical
topologies. In the present work we check the robustness of PSH in 
spherical topologies.

Let us consider various spherical spaces of the same order 120 --
namely the lens space $L(120,1)$, the binary dihedral space
$D^{*}_{30}$ and the Poincar\'e space $I^{*}$ -- and calculate the
corresponding PSH's.  The density parameters are fixed to
$\Omega_{m_{0}}=0.35$ and $\Omega_{\Lambda_0} = 0.75$.  In the runs,
the number of objects in the catalog is kept constant.  We examine
separately the effects of errors in position due to redshift
uncertainty $\Delta z$, and the effects of catalog incompleteness. 
Each of these effects will contribute to spoil the sharpness of the
topological signal.  For a given depth of the catalog, namely a
redshift cut-off $z_{cut}$, we perform the runs to look for the
critical value of the error at which the topological signal fades out.
 
\begin{figure}   
\centerline{  
\epsfig{figure=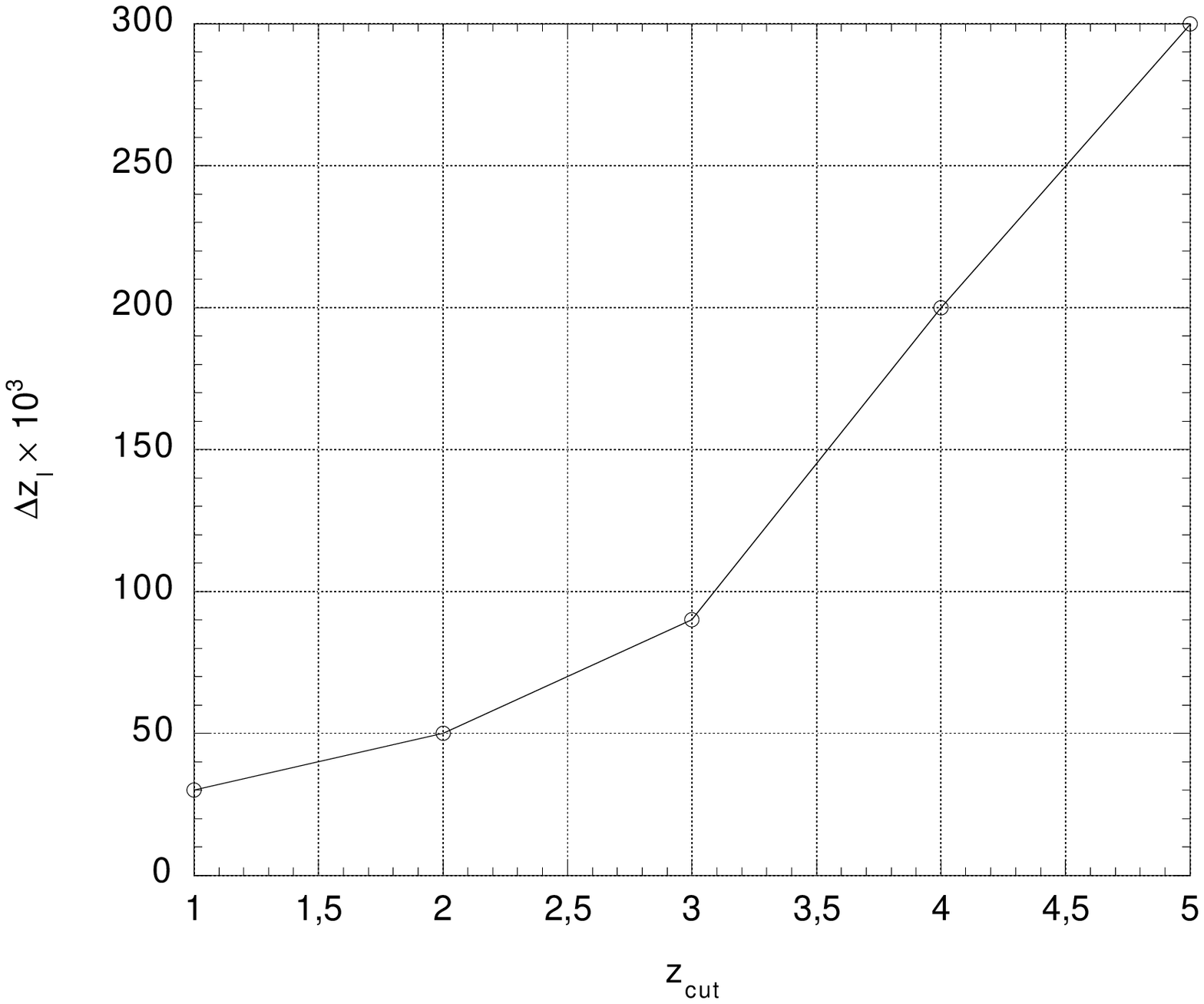,width=8cm}}   
\caption{Plot of $\Delta z_{l}$ as a function of the depth of the
catalog, for the spherical space $Z_{120}$ with $\Omega_0=0.35$,
$\Omega_{\Lambda0}=0.75$,
using the PSH method.}
\label{Getal_deltaz}   
\end{figure}

Figure \ref{Getal_deltaz} gives the critical redshift error $\Delta
z_{l}$ above which the topological spikes disappear for the spherical
space $Z_{120}$ with $\Omega_0=0.35$, $\Omega_{\Lambda0}=0.75$.
 
Next, we simulate an incomplete catalog where we randomly throw out
$p\%$ of the objects from the ideal catalog.  For the cyclic group
$Z_{120}$ and the binary dihedral group $D^{*}_{30}$, the topological
signal is destroyed only for a very large rejection percentage (above
90\%).  For the less favorable case of the Poincar\'e group $I^{*}$,
the results are summarized in table \ref{table5}.  Thus our
calculations confirm that the PSH method is perfectly robust for all
spherical topologies containing Clifford translations.

\begin{table}
\caption{Values of the limit percentage $p_{l}$ of rejection above which the  
{\sl PSH} spikes disappear, as a function of the redshift cut--off 
for the Poincar\'e space $I^{*}$.}   
\begin{center} 
\begin{tabular}{|c||c|c|c|} 
\hline 
$z_{\rm cut}$ & $\qquad1\qquad$ & $\qquad3\qquad$ & 
$\qquad\ge 5\qquad$  \\ 
\hline 
$p_{l}$ & $\qquad$ No signal$\qquad$ & $\qquad 80\%\qquad$ & 
$\qquad 90\%\qquad$\\ 
\hline 
\end{tabular} 
\end{center}
\label{table5}
\end{table}   

As we have seen the PSH method applies for most of the spherical
manifolds.  Nevertheless when the order of the group or of one of its
cyclic subgroups is too high then the spikes are numerous and have a
small amplitude.  Thus they may be very difficult to detect
individually.  This the case for instance for some $L(p,q)$ as well as
for linked action manifolds.  In that case the CCP method, which
gathers the topological signal into a single index, is more suitable. 
Again the topological signal is rather insensitive to reasonable
observational errors as soon as the underlying geometry contains
enough Clifford translations.
 
\section{Conclusion and Perspectives}
\label{VI} 

In this article, we have investigated the possible topologies of
a locally spherical universe in the framework of Friedmann--Lema\^{\i}tre
spacetimes. We have given the first primer of the classification
of three-dimensional spherical spaceforms, including the constructions
of these spaces.
  
We have determined the topologies which are likely to be detectable in
three--dimensional catalogs of cosmic objects using crystallographic
methods, as a function of the cosmological paramaters and the depth of
the survey.  The expected form of the Pair Separation Histogram is
predicted, including both the background distribution and the location
and height of the spikes.  We have performed computer simulations of
PSH in various spherical spaces to check our predictions.  The
stability of the method with respect to observational uncertainties in
real data was also proved.

Such a complete and exhaustive investigation of the geometrical 
properties of spherical spaces will be useful for further studies 
dealing with spherical topologies, including two--dimensional methods 
using CMB data. We plan to investigate the applicability of the
circle-matching method \cite{cornish98} for spherical topologies.

\appendix 
 
\section{Quaternions}\label{A} 
 
The quaternions are a four--dimensional generalization of the 
familiar complex numbers.  While the complex numbers have a single 
imaginary quantity $\bi$ satisfying $\bi^2 = -1$, the quaternions have three 
imaginary quantities $\bi$, $\bj$, and $\bk$ satisfying 
\begin{equation}\label{a1} 
\bi^2=\bj^2=\bk^2=-1 
\end{equation} 
which anti--commute 
\begin{equation}\label{a2} 
\lbrace\bi,\bj\rbrace=0,\quad 
\lbrace\bj,\bk\rbrace=0,\quad 
\lbrace\bk,\bi\rbrace=0 
\end{equation} 
and are subject to the multiplication rules 
\begin{equation} 
[\bi,\bj]=2\bk,\quad 
[\bj,\bk]=2\bi,\quad 
[\bk,\bi]=2\bj, 
\end{equation} 
and 
\begin{equation}\label{a2_1} 
[\bi,{\bf 1}]=0,\quad 
[\bj,{\bf 1}]=0,\quad 
[\bk,{\bf 1}]=0, 
\end{equation} 
where $[]$ and $\lbrace\rbrace$ are the usual commutation and 
anti--commutation symbols. 
 
Geometrically, the set of all quaternions  
\begin{equation} 
{\bf q}=a{\bf 1}+b\bi+c\bj+d\bk,\qquad (a,b,c,d)\in R^4 
\end{equation} 
defines four--dimensional Euclidean space, and the set of all unit length 
quaternions, that is, all quaternions $a{\bf 1} + b\bi + c\bj + d\bk$ 
satisfying $a^2 + b^2 + c^2 + d^2 = 1$, defines the 3--sphere. The 
quaternions are associative, $({\bf q}{\bf r}) {\bf s} = {\bf q} ({\bf 
r}{\bf s})$, even though they are not commutative. Introducing 
the conjugate quaternion $\bar{\bf q}$ of ${\bf q}$ by 
\begin{equation} 
\bar{\bf q}\equiv a{\bf 1}-b{\bi}-c{\bj}-d{\bk} 
\end{equation} 
and the modulus of ${\bf q}$ by 
\begin{equation} 
|{\bf q}|\equiv\sqrt{{\bf q}\bar{\bf q}}=\sqrt{a^2+b^2+c^2+d^2}  
\end{equation} 
a unit quaternion satisfies 
\begin{equation} 
|{\bf q}|=1\Longleftrightarrow\bar{\bf q}={\bf q}^{-1}. 
\end{equation} 
The identity quaternion ${\bf 1}$ is fundamentally different from the 
purely imaginary quaternions $\bi$, $\bj$ and $\bk$, but among 
the unit length purely imaginary quaternions $b\bi + c\bj + d\bk$ 
there is nothing special about the basis quaternions $\bi$, $\bj$ and $\bk$. 
Any other orthonormal basis of purely imaginary quaternions would 
serve equally well.
 
\noindent{\bf Lemma A1}: \emph{Quaternion change of basis}   
\vskip0.25cm 
\noindent Let $M$ be a $3 \times 3$ orthogonal matrix, and define 
\begin{eqnarray} 
\bi^\prime &=& M_{11} \bi  +  M_{12} \bj  +  M_{13} \bk  \nonumber\\ 
\bj^\prime &=& M_{21} \bi  +  M_{22} \bj  +  M_{23} \bk  \nonumber\\ 
\bk^\prime &=& M_{31} \bi  +  M_{32} \bj  +  M_{33} \bk  
\end{eqnarray} 
then $\bi^\prime$, $\bj^\prime$ and $\bk^\prime$ satisfy the quaternion 
relations 
\begin{eqnarray} 
\bi^{\prime 2} &=& \bj^{\prime 2} = \bk^{\prime 2} = -{\bf 1}  \nonumber\\ 
\bi^\prime\bj^\prime &=&\bk^\prime = -\bj^\prime\bi^\prime\nonumber\\ 
\bj^\prime\bk^\prime &=& \bi^\prime = -\bk^\prime\bj^\prime\nonumber\\ 
\bk^\prime\bi^\prime &=& \bj^\prime = -\bi^\prime\bk^\prime 
\end{eqnarray} 
 
Lemma A1 says that an arbitrary purely imaginary quaternion $b\bi + c\bj
+ d\bk$ may, by change of basis, be written as $b^\prime\bi^\prime$ .
If the purely imaginary quaternion $b\bi + c\bj + d\bk$ has unit length,
it may be written even more simply as $\bi^\prime$.  A
not--necessarily--imaginary quaternion $a{\bf 1} + b\bi + c\bj + d\bk$
may be transformed to $a^\prime{\bf 1} + b^\prime\bi^\prime$.  If it has
unit length it may be written as $\cos \theta ~ {\bf 1} + \sin \theta ~
\bi^\prime$ for some $\theta$.
 
The unit length quaternions, which we continue to visualize as the 3--sphere, 
may act on themselves by conjugation or by left or right 
multiplication.
 
\noindent{\bf Proposition A2}: \emph{Conjugation by quaternions}   
\vskip0.25cm 
\noindent Let ${\bf q}$ be a unit length quaternion.  According to the
preceding discussion, we may choose a basis $\lbrace {\bf 1},
\bi^\prime, \bj^\prime, \bk^\prime \rbrace$ such that ${\bf q} = \cos \theta
~ {\bf 1} + \sin \theta ~ \bi^\prime$ for some $\theta$.  It is easy
to compute how ${\bf q}$ acts by conjugation on the basis $\lbrace
{\bf 1}, \bi^\prime, \bj^\prime, \bk^\prime \rbrace$:
\begin{eqnarray} 
(\cos\theta\,{\bf 1}+\sin\theta\,\bi^\prime)\,{\bf 1}\,(\cos\theta\,{\bf1}-  
\sin\theta\,\bi^\prime) &=& {\bf 1} \\ 
(\cos\theta\,{\bf 1}+\sin\theta\,\bi^\prime)\,\bi^\prime\,(\cos\theta\,{\bf1}-  
\sin\theta\,\bi^\prime)&=& \bi^\prime \\ 
(\cos\theta\,{\bf 1}+\sin\theta\,\bi^\prime)\,\bj^\prime\,(\cos\theta\,{\bf 
1}-\sin \theta\,\bi^\prime) &=&  \cos2\theta\, 
\bj^\prime+\sin2\theta\,\bk^\prime\\ 
(\cos\theta\,{\bf 1}+\sin\theta\,\bi^\prime)\,\bk^\prime\,(\cos\theta\,{\bf 1} 
-\sin\theta\,\bi^\prime)&=-&\sin 2\theta\,\bj^\prime + \cos2\theta\,\bk^\prime
\end{eqnarray} 
 
Conjugation by any quaternion fixes ${\bf 1}$ (``the north pole''), so
the action is always confined to the ``equatorial 2--sphere'' spanned
by $\lbrace \bi^\prime, \bj^\prime, \bk^\prime \rbrace$.  Within the
equatorial 2--sphere, conjugation by the particular quaternion $\cos
\theta\,{\bf 1} + \sin\theta\,\bi^\prime$ rotates about the $\bi^\prime$
axis through an angle $2\theta$.  Unlike the preceding action by
conjugation, which always has fixed points, when the quaternions act
by left or right multiplication they never have fixed points.
 
\noindent{\bf Proposition A3}: \emph{Left multiplication by quaternions} 
\vskip0.25cm 
\noindent Let ${\bf q}$ be a unit length quaternion.  Choose a basis 
$\lbrace {\bf 1}, \bi^\prime, \bj^\prime, \bk^\prime \rbrace$ such that 
${\bf q} = \cos\theta\,{\bf 1}+\sin\theta\,\bi^\prime$ for some $\theta$. 
It is easy to compute how ${\bf q}$ acts by left multiplication 
on the basis $\lbrace {\bf 1}, \bi^\prime, \bj^\prime, \bk^\prime \rbrace$. 
\begin{eqnarray} 
(\cos \theta\,{\bf 1} + \sin \theta\,\bi^\prime) {\bf 1} =&&   
\cos \theta\,{\bf 1}  + \sin \theta\, \bi^\prime \\ 
(\cos \theta\,{\bf 1} + \sin \theta\,\bi^\prime) \bi^\prime =&-& 
 \sin \theta~  {\bf 1}  + \cos \theta\, \bi^\prime \\ 
(\cos \theta\,{\bf 1} + \sin \theta\,\bi^\prime) \bj^\prime =&& 
 \cos \theta~  \bj^\prime + \sin \theta\, \bk^\prime \\ 
(\cos \theta\,{\bf 1} + \sin \theta\,\bi^\prime) \bk^\prime =&-&\sin \theta 
\, \bj^\prime + \cos \theta\, \bk^\prime. 
\end{eqnarray} 
We see that left multiplication rotates ${\bf 1}$ towards $\bi^\prime$
while simultaneously rotating $\bj^\prime$ towards $\bk^\prime$.  The
result is a screw motion along so--called \emph{Clifford parallels}.
The Clifford parallels are geodesics, and they are homogeneous in the
sense that there is a parallel--preserving isometry of $S^3$ taking any
one of them to any other (see section~\ref{generalities}).  Action by
right multiplication is similar, but yields left--handed Clifford
parallels instead of right--handed ones.

\section{Matrices}\label{B} 

This appendix defines each finite subgroup of $SO(3)$
as an explicit set of rotations.

Lifting a subgroup of $SO(3)$ to the corresponding binary
subgroup of ${\cal S}^3$ is easy.
Proposition A.2, together with the change of basis principle
in Proposition A.1, implies that a rotation through an angle
$\theta$ about an axis $(x,y,z)$ is realized by both the unit
length quaternion
\begin{equation}
{\bf q}=  \cos\frac{\theta}{2} \, {\bf 1}
 + \frac{1}{\sqrt{x^2 + y^2 + z^2}}\sin\frac{\theta}{2} 
 \,\left(x\,\bi +y\,\bj +z\,\bk\right)\label{qq}
\end{equation}
and by its negative.  This one-to-two mapping yields a binary
subgroup of ${\cal S}^3$ that is exactly twice as big as the
original subgroup of $SO(3)$.

Converting from a quaternion in ${\cal S}^3$ to a matrix in $SO(4)$ is
also easy.  When a quaternion ${\bf q}=a{\bf 1}+b\bi+c\bj+d\bk$ acts on
the left its effect on the basis vectors ${\bf 1}$, $\bi$, $\bj$, and
$\bk$ is, respectively,
\begin{eqnarray} 
(a{\bf 1} + b\bi + c\bj + d\bk) \,{\bf 1} &= & a{\bf 1} + b\bi +c\bj +d\bk \nonumber\\
(a{\bf 1} + b\bi + c\bj + d\bk) \, \bi    &=-& b{\bf 1} + a\bi +d\bj -c\bk \nonumber\\
(a{\bf 1} + b\bi + c\bj + d\bk) \, \bj    &=-& c{\bf 1} - d\bi +a\bj +b\bk \nonumber\\
(a{\bf 1} + b\bi + c\bj + d\bk) \, \bk    &=-& d{\bf 1} + c\bi -b\bj +a\bk
\end{eqnarray} 
Thus, relative to the orthonormal basis $\lbrace {\bf 1}, \bi, \bj, \bk
\rbrace$, the matrix for this action is

\begin{equation}
M_{\rm left}({\bf q})=\left( 
\begin{array}{rrrr} 
      a  & -b  & -c  & -d  \\ 
      b  &  a  & -d  &  c  \\ 
      c  &  d  &  a  & -b  \\ 
      d  & -c  &  b  &  a  \\ 
\end{array}\right).
\end{equation}
Similarly, when the same quaternion ${\bf q}=a{\bf 1}+b\bi+c\bj+d\bk$
acts on the right, the resulting isometry has matrix
\begin{equation}
M_{\rm right}({\bf q})=
\left( 
\begin{array}{rrrr} 
      a  & -b  & -c  & -d  \\ 
      b  &  a  &  d  & -c  \\ 
      c  & -d  &  a  &  b  \\ 
      d  &  c  & -b  &  a  \\ 
\end{array}\right). 
\end{equation}

In the case of double action and linked action manifolds
(\S\ref{da} and \S\ref{la}), one quaternion $r$ acts on the left
(giving a right--handed Clifford translation) while
a different quaternion $l$ acts on the right
(giving a left--handed Clifford translation).
Their combined action is given by the product of their
respective matrices.  The order in which we multiply
the two matrices doesn't matter because right-- and left--handed
Clifford translations always commute.

The following realizations of the finite subgroups of $SO(3)$
are unique up to an orthonormal change of coordinates.
 
\subsection{Cyclic Groups $Z_n$} 

Each cyclic group $Z_n$ consists of rotations about
the axis (0,0,1) through an angle $2 \pi k / n$, for $0 \leq k < n$.
This defines $n$ distinct rotations. 
 
\subsection{Dihedral Groups $D_m$} 

The dihedral group $D_m$ contains the $m$ rotations of the
cyclic group $Z_m$ about the axis (0,0,1)
as well as $m$ half turns about the axes
$( \cos \frac{k\pi}{m}, \sin \frac{k\pi}{m}, 0)$ 
for $0 \leq k < m$.
This defines $2m$ distinct rotations. 
 
\subsection{The Tetrahedral Group $T$} 
 
\begin{figure}[ht] 
\centerline{\epsfig{file=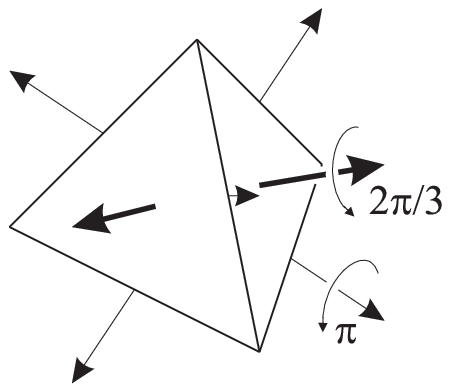, width=6cm}} 
\caption{Visualisation of the symmetry group $T$ of the tetrahedron.} 
\label{t1} 
\end{figure}

The tetrahedral group (Figure~\ref{t1}) consists of 
\begin{itemize}
    \item
        the identity,
    \item
        order 2 rotations about the midpoints of
        the tetrahedron's edges, realized as $\pi$ rotations
        about the three axes (1,0,0), (0,1,0), and (0,0,1), and
    \item
        order 3 rotations about the tetrahedron's vertices
        and face centers, realized as counterclockwise $2\pi/3$
        rotations about the eight axes $(\pm 1, \pm 1, \pm 1)$.
\end{itemize} 
This defines a total of $1+3+8=12=|T|$ distinct rotations.  Applying
equation (\ref{qq}) yields the 24 quaternions of the binary
tetrahedral group $T^\ast$.

\subsection{The Octahedral Group $O$} 

The octahedral group consists of
\begin{itemize}
    \item
        the identity,
    \item
        order 2 rotations about the midpoints of
        the octahedron's edges, realized as $\pi$ rotations
        about the six axes $(\pm 1, 1, 0)$,
        $(0, \pm 1, 1)$, and $(1, 0, \pm 1)$, and
    \item
        order 2 rotations about the octahedron's vertices,
        realized as $\pi$ rotations about the three axes
        (1,0,0), (0,1,0), and (0,0,1), and
    \item
        order 3 rotations about the centers of the octahedron's
        faces, realized as counterclockwise $2\pi/3$
        rotations about the eight axes $(\pm 1, \pm 1, \pm 1)$.
    \item
        order 4 rotations about the octahedron's vertices,
        realized as counterclockwise $\pi/2$
        rotations about the six axes
        $(\pm 1,0,0)$, $(0,\pm 1,0)$, and $(0,0,\pm 1)$.
\end{itemize}
This defines a total of $1+6+3+8+6=24=|O|$ distinct rotations.
Applying equation (\ref{qq}) yields the 48 quaternions of the binary
octahedral group $O^\ast$.  Note that the octahedral group contains the
tetrahedral group within it.
 
\subsection{Icosahedral Group $I$} 

The icosahedron's twelve vertices being at
$(\pm \phi, \pm 1, 0)$, $(0, \pm \phi, \pm 1)$, and
$(\pm 1, 0, \pm \phi)$, where $\phi = (\sqrt{5} - 1)/2$
is the golden ratio, the icosahedral group consists of
\begin{itemize}
    \item the identity,
    \item order 2 rotations about the midpoints of
        the icosahedron's edges, realized as $\pi$ rotations
        about the fifteen axes (1,0,0), (0,1,0), (0,0,1),
        $(\pm \phi, \pm (\phi + 1), 1)$,
        $(1, \pm \phi, \pm (\phi + 1))$, and
        $(\pm (\phi + 1), 1, \pm \phi)$, and
    \item order 3 rotations about the centers of the icosahedron's
        faces, realized as counterclockwise $2\pi/3$
        rotations about the twenty axes $(\pm 1, \pm 1, \pm 1)$,
        $(0, \pm (\phi + 2), \pm 1)$,
        $(\pm 1, 0, \pm (\phi + 2))$,
        $(\pm (\phi + 2), \pm 1, 0)$, and
    \item order 5 rotations about the icosahedron's vertices,
        realized as both counterclockwise $2\pi/5$
        rotations and counterclockwise $4\pi/5$ rotations
        about the twelve axes $(\pm \phi, \pm 1, 0)$,
        $(0, \pm \phi, \pm 1)$, and $(\pm 1, 0, \pm \phi)$.
\end{itemize}

This defines a total of $1+15+20+24=60=|I|$ distinct rotations.
Applying equation (\ref{qq}) yields the 120 quaternions of the binary
icosahedral group $I^\ast$.  Note that the icosahedral group also contains
the tetrahedral group within it.

\ack{JRW thanks the MacArthur Foundation
for its support, and thanks the remaining authors for their
warm hospitality. E.G. thanks FAPESP--Brazil (Proc. 00/04911-8)
for financial support.}\vskip 0.5cm


\begin{thebibliography}{30} 
 
\bibitem{sndata} 
A.G. Riess {\em et al.}, Astron.  J. {\bf116} (1998) 1009;
S.Perlmutter {\em et al.}, Nature (London) {\bf 391} (1998) 51.
 
\bibitem{cmbdata} 
P. de Bernardis {\em et al.}, Nature (London) {\bf404} (2000) 955;
{\it ibid}, [{\tt astro-ph/0105296}].

\bibitem{gldata} 
Y. Mellier,  Ann. Rev. Astron. Astrophys. {\bf37} (1999) 127. 
 
\bibitem{vdata} 
R. Juszkiewicz {\em et al.},  Science {\bf 287} (2000) 109. 
 
\bibitem{srdata} 
B. Roukema and G. Mamon,  Astron. Astrophys. {\bf 358} (2000) 395.

\bibitem{quint}
C. Wetterich, Nucl. Phys. {\bf B302} (1988) 668;
R.R. Caldwell, R. Dave, and P.J. Steinhardt, 
Phys. Rev. Lett. {\bf80} (1998) 1582;
I. Zlatev, L. Wang, and P.J. Steinhardt, Phys. Rev. Lett. {\bf82} (1999) 896;
P.G. Ferreira and M. Joyce, Phys, Rev. {\bf D58} (1998) 023503;
J.--P. Uzan, Phys. Rev. {\bf D59} (1999) 123510.

\bibitem{wsp}  
M. White, D. Scott, and E. Pierpaoli,  {\tt [astro-ph/0004385]}. 
 
\bibitem{jaf}   
A. H. Jaffe {\em et al.},  Phys.  Rev. Lett. {\bf 86} (2000) 3475.
 
 \bibitem{Mos73}  
G.D. Mostow, Ann. Math. Studies {\bf 78} (1973), Princeton 
University Press, Princeton, New Jersey. 
 
\bibitem{lachieze95}   
M.~Lachi\`eze-Rey and J.-P.~Luminet,  
 Phys.  Rep.  {\bf 254} (1995) 135.  
 
\bibitem{uzan97} 
J.--P. Uzan, Int. J. Theor. Physics {\bf36} (1997) 2439. 
 
\bibitem{luminet99}  
J-P. Luminet and B.F. Roukema, in {\it Theoretical and   
Observational Cosmology}, M.~Lachi\`eze-Rey (Ed.), Kluwer Ac. Pub.,   
pp.117-157, [{\tt astro-ph/9901364}]. 
 
\bibitem{uzan99b}   
J-P. Uzan, R.~Lehoucq, and J-P.~Luminet, 
Proc. of the XIX$^{\rm th}$ Texas meeting, Paris 14--18 December 1998,  
Eds. E. Aubourg, T. Montmerle, J. Paul and P. Peter, article n$^{\rm o}$  
04/25, [{\tt gr-qc/0005128}]. 
 
\bibitem{lehoucq96}   
R.~Lehoucq, M.~Lachi\`eze-Rey, and J-P.~Luminet,  
Astron.  Astrophys. {\bf 313} (1996) 339. 
 
\bibitem{lehoucq99}   
R.~Lehoucq, J-P.~Luminet, and J-P. Uzan,  
Astron. Astrophys. {\bf 344} (1999) 735. 

\bibitem{gomero98}  
G.I. Gomero, A.F.F. Texeira, M.J. Rebou\c{c}as, and 
A. Bernui, 
[{\tt gr-qc/9811038]}. 

\bibitem{fagundes98}   
H.V. Fagundes and E. Gausmann,  
[{\tt astro-ph/9811368}]. 
 
\bibitem{fagundes99}  
H.V. Fagundes and E. Gausmann, Phys. Lett. {\bf A261} (1999) 235. 
 
\bibitem{gomero99}    
G.I. Gomero, M.J. Rebou\c{c}as, and A.F.F. Teixeira, [{\tt 
astro-ph/9909078}]; {\it ibid.}, [{\tt astro-ph/9911049}]. 
 

\bibitem{lehoucq00}  
R.~Lehoucq, J-P. Uzan, and J-P.~Luminet,  
Astron. Astrophys. {\bf363} (2000) 1. 
 
\bibitem{uzan99}   
J-P. Uzan, R.~Lehoucq, and J-P.~Luminet,  
Astron.  Astrophys. {\bf 351}  (1999) 766. 

\bibitem{wolf83}  
J.A. Wolf, {\it 
Spaces of constant curvature}, fifth edition, Publish or 
Perish Inc., Wilmington USA (1967). 
 
 \bibitem{beardon83}  
A.F. Beardon, 
{\it The geometry of discrete groups}, New York, Springer (1983). 
 
\bibitem{nakahara90}  
M. Nakahara, {\it Geometry, Topology and 
Physics}, Adam Hilger, Bristol, New-York (1990).

\bibitem{deS17}  
W. de Sitter, Month. Not. R. Astron. Soc {\bf 78} (1917) 3 ;  Proceedings 
of the Royal Academy of Amsterdam {\bf 20} (1917) 229. 
 
\bibitem{Lum97}  
J.-P. Luminet, {\sl L'invention du big bang}, introduction to 
{\sl Essais de 
Cosmologie de A. Friedmann et G. Lema\^\i tre}, Seuil, Paris, 1997, 
p. 78.  

\bibitem{Edd23} 
 A.S. Eddington, {\sl The mathematical theory of 
relativity}, chap. 5, (Cambridge University Press, Cambridge, 1923). 
 
\bibitem{Fri24}  
A. Friedmann, Zeitschr. f\"ur Phys. {\bf 2} (1924) 326. 
 
\bibitem{Lem29}  
G. Lema\^\i tre, Rev. Quest. Sci. (1929) 189 
(English translation in {\sl The Primeval atom},   
Van Nostrand, New York, 1950). 
 
\bibitem{Nar85}  
J.V. Narlikar and T.R. Seshadri, Astrophys. J. {\bf 288} (1985) 43. 
 
\bibitem{Ell71}
G.F. R. Ellis, Gen. Rel. Grav. {\bf 2} (1971) 7.

\bibitem{Got80}  
J. R. Gott, Month.  Not.  R. Astron. Soc. {\bf 193} (1980) 153. 
 
\bibitem{Ioni98} 
R. Ionicioiu and R. Williams, Class. Quant. Grav. {\bf 15} (1998) 3469.  

\bibitem{gom01}    
G.I. Gomero, M.J. Rebou\c{c}as, and R. Tavakol,  
[{\tt gr-qc/0105002}].

\bibitem{ThrelfallSeifert}   
W. Threlfall and H. Seifert,  
``Topologische Untersuchung der 
Diskontinuit\"atsbereiche endlicher Bewegungsgruppen des 
dreidimensionalen sph\"arischen Raumes'', 
Math. Annalen {\bf 104} (1930) 1--70 and {\bf 107} (1932) 543--586.

\bibitem{thurston97}  
W.P. Thurston,  
{\it Three-dimensional geometry  
and topology} (1997) Princeton Mathematical series {\bf35}, Ed.  S.  
Levy, (Princeton University Press, Princeton, USA). 
 
\bibitem{Poincare04}   
H. Poincar\'e,  
``Cinqui\`eme compl\'ement \`a l'analysis situs'', 
Rend. Circ. Mat. Palermo {\bf 18} (1904) 45-110. 
 
\bibitem{WeberSeifert33}   
C. Weber and H. Seifert,  
``Die beiden Dodekaederr\"aume'', 
Math. Zeitschrift {\bf 37} (1933) 237-253. 

\bibitem{bernui} A. Bernui and A.F.F. Teixeira, [{\tt
astro-ph/9904180}].
 
\bibitem{cornish98}   
N. Cornish, D. Spergel, and G. Starkman, Class.  Quant.  Grav.  {\bf
15} (1998) 2657.


\end{thebibliography}
\end{document}